\documentclass[a4paper,11pt,preprint]{article}
\pdfoutput=1

\usepackage{jcappub}
\usepackage{multirow}
\usepackage{graphicx,color}% Include figure files
\usepackage{dcolumn}% Align table columns on decimal point
\usepackage{bm}% bold math
\usepackage{enumerate}
\usepackage{color}
\usepackage{epsfig}
\usepackage{hyperref}
\usepackage{natbib}
\usepackage{float}
\usepackage{array}
\usepackage{ragged2e}
\usepackage{adjustbox}
\usepackage{dsfont}
\usepackage{epstopdf}
\usepackage{amssymb,amsmath,amsthm,mathrsfs}% outlined letters
\usepackage{soul}

\newcommand{\overbar}[1]{\mkern 1.5mu\overline{\mkern-1.5mu#1\mkern-1.5mu}\mkern 1.5mu}

\makeatletter
\newcommand\xleftrightarrow[2][]{%
	\ext@arrow 9999{\longleftrightarrowfill@}{#1}{#2}}
\newcommand\longleftrightarrowfill@{%
	\arrowfill@\leftarrow\relbar\rightarrow}
\makeatother

\def\be{\begin{equation}}
\def\ee{\end{equation}}
\def\ben{\begin{equation*}}
\def\een{\end{equation*}}
\def\bea{\begin{eqnarray}}
\def\eea{\end{eqnarray}}
\def\bal{\begin{align}}
\def\eal{\end{align}}

\newcommand{\nn}{\nonumber \\}

\title{Inconsistency of an inflationary sector coupled only to Einstein gravity}

\newcommand{\addressEPFL}{Institute of Physics, Laboratory of Particle Physics and Cosmology (LPPC), \'Ecole Polytechnique F\'ed\'erale de Lausanne (EPFL), CH-1015 Lausanne, Switzerland.}
\author{Daniel G. Figueroa}
\author{and Erwin H. Tanin}
\affiliation{\addressEPFL}

\emailAdd{daniel.figueroa@epfl.ch}
\emailAdd{ehtanin@gmail.com}
\date{\today}

\abstract{From a model-building perspective, the inflationary sector might very well have no direct couplings to other species, apart from inevitable gravitational interactions. Within the context of General Relativity, a thermal universe can still emerge after inflation if: $i)$ some radiation sector is excited towards the end of inflation, and $ii)$ the post-inflationary equation of state becomes sufficiently stiff $w \geq w_{\rm RD}\gtrsim 0.57$, with $w_{\rm RD}$ a threshold depending on the inflationary scale $H_*$ and the initial radiation-to-inflaton energy ratio $\Delta_*$. Furthermore, a stiff period in the expansion history enhances significantly the inflationary gravitational wave (GW) background, making this signal (potentially) observable by aLIGO, LISA and other experiments. The very same enhancement leads however to an inconsistency of the scenario: the energy of the GWs becomes too large compared to the rest of the radiation sector, violating standard BBN and CMB bounds on GW backgrounds. Except for very special scenarios where the initial radiation sector comprises hundreds of fields with couplings tuned to specific values, our result applies independently of $w$, $H_*$ and $\Delta_*$. This suggests that in order to reheat the universe, the inflationary sector should be coupled directly to other particle species. Alternatively the inflationary sector could be implemented in modified gravity theories. }

\begin{document}
\maketitle

\section{Introduction}

Compelling evidences support the idea that the Universe underwent an early phase of accelerated expansion, $inflation$~\cite{Akrami:2018odb}. As the specific particle physics `embedding' of inflation is unknown, inflation is typically described in terms of a singlet scalar field, the $inflaton$, with potential and initial conditions appropriately chosen. In order to switch to the standard hot Big Bang cosmology after inflation, the inflaton energy must be converted into radiation degrees of freedom ($dof$), which eventually thermalize and dominate the Universe's energy budget. This transition process is known as $reheating$. 

Though the reheating process has not yet been probed observationally, we know that the radiation domination (RD) epoch must begin before the onset of Big Bang Nucleosynthesis (BBN). In general, the details of reheating depend on the choice of inflationary model (e.g.~the inflaton potential) and on the inflaton couplings to other matter sectors. Thus, in a given scenario, one typically invokes the need for having such coupling(s) and even determine some of their properties (e.g.~their strength) based on the need to reheat successfully the Universe before BBN.

If the inflaton has sufficiently large couplings to other species, the initial stages following after inflation are typically characterized by non-perturbative particle production phenomena, which dominate over inflaton perturbative decays. This is known as the $preheating$ stage. Various mechanisms of preheating, tailored for different inflationary models and coupling constructions, have been investigated in detail in the past, from parametric resonance~\cite{Shtanov:1994ce,Kofman:1997yn,Figueroa:2016wxr}, spinodal instabilities~\cite{Felder:2000hj,Felder:2001kt} and instant preheating~\cite{Felder:1998vq,Felder:1999pv} of scalar fields, to parametric excitation of fermions~\cite{Greene:1998nh,Greene:2000ew,Giudice:1999fb} and non-perturbative excitation mechanisms of gauge fields~\cite{Dufaux:2010cf,Deskins:2013lfx,Adshead:2015pva,Adshead:2017xll,Tranberg:2017lrx}. For reviews on preheating and exhaustive lists of references, see~\cite{Allahverdi:2010xz,Amin:2014eta}. 

A reheating mechanism that deserves a class of its own is $gravitational$ $reheating$. While the term ``gravitational reheating" may have a broader meaning in the context of modified gravity, here we use it to refer to the specific reheating model originally proposed in~\cite{Ford:1986sy} (see also~\cite{Spokoiny:1993kt}). In this construction the gravitational sector is described by General Relativity (GR), and the inflaton is assumed to have no direct coupling to any matter sector, apart from inevitable gravitational interactions. Two ingredients are needed for this scenario to work: $i)$ some light $dof$ need to be excited either during inflation or towards the end of it, due to their gravitational coupling, and $ii)$ the equation of state (EoS) $w$ of the inflaton must become stiff subsequently after inflation, i.e.~$1/3 < w \leq 1$. Actually, in order reheat before BBN, the EoS must become at least as large as $w \gtrsim 0.57$, as we will show. In general, by the end of inflation, the energy density of the radiation produced gravitationally is typically too small compared to the energy density of the inflaton. Nevertheless, thanks to requirement $ii)$, the energy density of the inflaton will decrease faster with the expansion of the universe than the initially subdominant energy density of the radiation sector. Hence, eventually the radiation will become the dominant energy component of the Universe, and once it has thermalized, the desired initial condition for the hot Big Bang thermal era is obtained: a universe dominated by relativistic species in thermal equilibrium. Different implementations of an early stiff era and its cosmological consequences (related or not to gravitational reheating), can be found e.g.~in~\cite{Joyce:1996cp,Joyce:1997fc,Peebles:1998qn,Sahni:2001qp,Liddle:2003zw}, or more recently in~\cite{Chavanis:2014lra,Figueroa:2016dsc,Li:2016mmc,Dimopoulos:2018wfg,Haro:2018jtb}.

During inflation, quantum fluctuations inevitably give rise to a (quasi-)scale-invariant spectrum of tensor metric perturbations at super-Hubble scales. During the evolution of the Universe after inflation, when the tensor modes cross inside the Hubble radius during the RD stage, they become a stochastic background of gravitational waves (GWs) with a (quasi-)scale invariant energy density spectrum. It is however well known that if there is an early stiff era before RD, as required in gravitational reheating, the (quasi-)scale invariance is broken and the GW spectrum becomes significantly blue tilted in the frequency range corresponding to the modes crossing the horizon during the stiff period~\cite{Giovannini:1998bp,Giovannini:1999bh,Riazuelo:2000fc,Sahni:2001qp,Tashiro:2003qp,Boyle:2007zx,Giovannini:2008zg,Giovannini:2008tm,Artymowski:2017pua,Caprini:2018mtu}. In this case, the amplitude of the inflationary tensor power spectrum becomes considerably enhanced, relative to the amplitude in the standard case where the universe is assumed to become RD right after the end of inflation. The origin of this enhancement can be understood simply as follows (see Appendix~\ref{appendix:GWspectrumderivation} and \cite{Figueroa:2019paj} for detailed derivations). By definition, the GW energy density spectrum $\Omega_{\rm GW}$ is defined as the GW energy density $\rho_{\rm GW}$ per unit logarithmic wavenumber $k$ interval, normalized to the critical density $\rho_{\rm crit}$. During the stiff epoch following the end of inflation, characterized by an equation of state $w$, the critical density scales as $\rho_{\rm crit}\propto a^{-3(1+w)}$. As sub-horizon GWs scale as radiation $\rho_{\rm GW}\propto a^{-4}$, the GW energy spectrum must scale then as $\Omega_{\rm GW}\equiv\frac{1}{\rho_{\rm crit}}\frac{d\rho_{\rm GW}}{d\log k}\propto a^{3w-1}$. By the end of the stiff epoch (beginning of RD), each mode has scaled an amount $\Omega_{\rm GW}(k)\propto (a_{\rm RD}/a(t_k))^{3w-1}$, where $t_k$ is the conformal time when a given mode $k$ crossed the horizon. Now, a mode with wavenumber $k$ crosses the horizon when $kt_k\sim 1$, whereas the Friedmann equation says that $a(t_k)\propto t_k^{2/(1+3w)}$. The resulting GW spectrum is therefore $\Omega_{\rm GW}\propto k^{2(3w-1)/(1+3w)}$, which is blue-tilted for a stiff equation of state $w > 1/3$. In the subsequent epochs, all the modes that crossed the horizon during the stiff epoch experience the same amount of redshifting, and consequently the shape of the blue-tilted part of the GW energy spectrum remains unchanged.

In this paper, we will show that standard implementation of gravitational reheating is actually inconsistent, as the mentioned blue-tilted GW spectrum due to a stiff era, is in contradiction with basic cosmological constraints. The logic is the following. As it is well known, in order for BBN to proceed successfully, the energy budget of the Universe must not comprise a significant amount of extra relativistic species, including GWs. This condition, known as the BBN bound on stochastic GW backgrounds, requires that the energy fraction of GWs to Standard Model (SM) radiation $dof$ in that moment, must not be larger than $\sim 10\%$~\cite{Caprini:2018mtu}. We will arrive however at the striking result that when the Universe becomes RD at the end of the stiff era (before BBN), the energy budget of the Universe is dominated by GWs, and not by the radiation component excited towards the end of inflation, in clear contradiction with the above bound. The problem becomes even more severe when considering the upper bound on stochastic GW backgrounds from the Cosmic Microwave Background (CMB), which requires the fraction of energy density of GWs to SM radiation $dof$ to be no larger than $\sim 1\%$~\cite{Smith:2006nk,Sendra:2012wh,Caprini:2018mtu}. As we will show, our result is independent of the parameter space, i.e.~independent of the inflationary Hubble rate and of the value of the stiff EoS after inflation. 

From now on, {\small$m_p = {1\over\sqrt{8\pi G}} \simeq 2.44\cdot10^{18}$ GeV} is the reduced Planck mass, {\small$a(t)$} is the scale factor, {\small$t$} is the conformal time, and we use Friedman-Lema\^itre-Roberson-Walker (FLRW) metric $ds^2 = a^2(t)\eta_{\mu\nu}dx^\mu dx^\nu$. A subscript $*$ denotes evaluation at the end of inflation.

\section{Gravitational reheating} 
\label{sec:GravRHidea}

Let us consider a scalar field inflaton $\phi$ which does not interact with any other matter field, unless gravitationally. We simply demand that the inflaton potential $V(\phi)$ is engineered to sustain a (quasi-)$de~Sitter$ stage of inflation in the early Universe, but we will not specify the form of the potential that achieves this goal. Since after inflation the expansion of the universe must be eventually driven by a plasma of relativistic species with large thermal energy density, such particles must somehow be excited out of the energy available in the inflationary sector. As we will see, such excitation is possible because of their gravitational coupling to the inflationary sector. However, it is well known that this process is highly inefficient. The energy density of the radiation that can be produced in this way is typically much less than the energy stored in the inflationary sector at the end of inflation. In order to reheat successfully, this energy hierarchy needs to be inverted, meaning that a mechanism for amplifying the energy density fraction of the produced radiation is required. In the gravitational reheating scenario, where inflation is immediately followed by a stiff epoch, the amplification occurs naturally as the energy density of radiation scales down more slowly than that of a stiff fluid. Thus, it is merely a matter of time that the energy hierarchy will be inverted.

\subsection{Initial particle production}
\label{subsec:PartProd}

The radiation sector may, in principle, comprise of fields of any spin. That being said, it is well known that only massless or sufficiently light scalar (spin 0) and tensor (spin 2) $dof$ are amplified during inflation\footnote{Here were are assuming that gravity is described by general relativity. Metric vector perturbations are actually excited as well during inflation, but, unlike scalar and tensor $dof$, they decay promptly.} due to their gravitational couplings to the (quasi-) $de~Sitter$ background. Other $dof$ such as massless fermions, gravitinos, and gauge fields do not get excited, as they are conformally invariant~\cite{Birrell:1982ix}. 

Scalar fields non-minimally coupled to gravity (with non-conformal coupling) can also be excited towards the end of inflation, if the background curvature changes sufficiently fast. Using standard techniques of quantum field theory in time dependent backgrounds, as originally discussed in~\cite{Ford:1986sy,Spokoiny:1993kt}, it can be shown that a scalar field with a non-minimal coupling to gravity of the form ${1\over2}\xi\varphi^2R$, can be largely excited towards the end of inflation. This happens when the curvature {\small$R$} changes sufficiently fast (non-adiabatically) during the transition from inflation to the power-law regime following afterwards. Ford~\cite{Ford:1986sy} calculated the energy density of the created species in the limit of nearly conformal coupling $|1-6\xi| \ll 1$, and found
\begin{align}\label{eq:almostconf}
\rho_{\rm rad} \sim 10^{-2}(1-6\xi)^2H_*^4\,,
\end{align}
where $H_*$ is the Hubble rate at the end of inflation, and the prefactor $(1-6\xi)^2$ [implicitly assumed to be small $\ll 1$] correctly exhibits the absence of particle creation if the coupling is conformal $\xi = {1\over 6}$. This result was later generalized to other coupling values satisfying $|1-6\xi| \lesssim 1$, obtaining~\cite{Damour:1995pd}
\begin{align}\label{eq:almostarbitrary}
\rho_{\rm rad} \sim 10^{-2}H_*^4\,,
\end{align}
which naturally replaces the factor $(1-6\xi)^2$ present in the nearly conformal case by some value of order unity. If $\xi \gg 1$, however, the energy of the excited species is suppressed by the large coupling~\cite{Herranen:2015ima}, and hence we will not consider this case from now on. 

In the case where there are several species, say $\mathcal{N}_{1}$ fields with non-minimal couplings restrained to $|6\xi-1| \sim 1$,  and $\mathcal{N}_{2}$ almost conformally coupled fields with $|1-6\xi| \ll 1$, the total amount of energy stored in all these fields at the end of inflation, would read
\begin{align}\label{eq:totalNonMinFlds}
\rho_{\rm rad} \sim 10^{-2}H_*^4\left(\mathcal{N}_{1} + \sum_{i=1}^{\mathcal{N}_{2}}(1-6\xi_i)^2 \right)\,.
\end{align}
If $\mathcal{N}_{1} \gg 1$ then most likely the contribution from the first term will dominate over the second contribution. However, requiring such a large number of fields to have non-minimal couplings tuned to $|6\xi-1| \sim 1$ seems unappealing, as we will comment further in Section~\ref{sec:MinimalRequisites}.

In addition to non-adiabatic production of particles at the end of inflation, spectator scalar fields with potential $V = {1\over2}m^2\varphi^2$ and mass $m^2 \ll H_*^2$, i.e.~free light scalar $dof$, can also be excited during inflation out of initially sub-Hubble quantum fluctuations. By the end of inflation, these fields would have accumulated an energy density of the order of
\begin{align}\label{eq:FreeFields}
\rho_{\rm rad} \sim 10^{-2}H_*^4\times \Delta N\sum_{i}^{\mathcal{N}_f}\left(m_i\over H_*\right)^2\,,
\end{align}
where $\Delta N$ represents the total number of e-folds during inflation, $\mathcal{N}_f$ the total number of these spectator fields, and $m_i$ their masses. In general, unless the masses are tuned to $1/\sqrt{\Delta N} \lesssim m_i/H_* \lesssim 1$, we expect the factor $\Delta N\sum_{i}^{\mathcal{N}_f}(m_i/H_*)^2$ to be $\ll 1$, modulo a possible enhancement linear in the number of fields. It is however not particularly motivated to assume the existence of many such free fields with a large mass, completely decoupled from other  $dof$. Hence, we rather expect in general to have $\rho_{\rm rad}\ll 10^{-2}H_*^4$. In the presence of a self-interaction potential $V = {\lambda\over4}\varphi^4$, the fields would still accumulate a total energy density by the end of inflation
\begin{align}\label{eq:SelfInteracting}
\rho_{\rm rad} \sim \mathcal{N}_{\rm si}\times10^{-2}H_*^4\,,
\end{align}
where $\mathcal{N}_{\rm si}$ is the total number of such fields. A natural example of such a self-interacting spectator field (and actually one that we know for certain to exist) is the Higgs field of the Standard Model~\cite{Enqvist:2013kaa,Figueroa:2015rqa,Espinosa:2015qea,Figueroa:2016dsc}. Notice that Eq.~(\ref{eq:SelfInteracting}) is actually independent of the self-couplings $\lambda$ of these fields. The reason is that during inflation the self-interaction of each of these fields provides a dynamical mass $m_{\varphi}^2 \sim \lambda\langle\varphi^2\rangle$. The mass would grow until an equilibrium distribution of the field fluctuations is achieved, that is, when the condition $\lambda\varphi^4 \sim H_*^4$ is satisfied~\cite{Starobinsky:1994bd}. Then the distribution is preserved during the remaining e-folds of inflation, and hence the energy density of the field is $V \sim \lambda \varphi^4 \sim H_*^4$, which is independent of $\lambda$.

In summary, whenever light scalar $dof$ that are free, self-interacting, and/or non-minimally coupled to gravity, exist during inflation, it is natural to expect that they will be excited with a total energy density at the end of inflation, which can be parametrized as 
\begin{align}\label{eq:EnergyParam}
\rho_{\rm rad} = \delta \times 10^{-2}H_*^4\,,
\end{align}
with $\delta$ a dimension-less number characterizing the specific details of a given case. Based on the preceding discussions, we typically expect $\delta \ll 1$ or at most $\delta \lesssim 1$. As we will explain further in Sect.~\ref{sec:MinimalRequisites}, even though it is possible to conceive scenarios with $\delta \gg 1$,  they correspond to $ad~hoc$ constructions requiring the presence of hundreds of fields during inflation, with couplings appropriately tuned. We will therefore take $\delta\leq 1$ as the canonical/generic range.%, understanding that typically we expect $\delta \ll 1$ or $\delta \lesssim 1$. %whereas $\delta \gg 1$ represents a rather contrived situation.

It is remarkable that even though the energy density in Eq.~(\ref{eq:EnergyParam}) can be rather large for high energy models of inflation, it is still very subdominant compared to the energy density in the inflationary sector at the end of inflation, 
\begin{align}\label{eq:InitialRatio}
\Delta_* \equiv {\rho_{\rm rad} \over 3 m_p^2 H_*^2} = {\delta\over 300}\left({H_*\over m_p}\right)^2 \sim \delta\cdot 10^{-12}\times\left({H_*\over H_{\rm max}}\right)^2~\ll~ 1\,
\end{align}
where we have normalized $H_*$ to the current upper bound on the inflationary scale
\begin{align}\label{eq:Hmax}
H_{\rm max} \simeq 6.6\cdot 10^{13}\,{\rm GeV}\,,
\end{align}
as set by the most recent measurement of the B-mode polarization anisotropies of the CMB~\cite{Akrami:2018odb,Ade:2018gkx}, see discussion below Eq.~(\ref{eq:consistencyCondition}). Let us note that Eq.~(\ref{eq:Hmax}) indicates the energy scale when the CMB scales left the Hubble radius during inflation, 50-60 e-foldings before the end of inflation. In the case of slow-roll inflation, the Hubble rate during inflation is not constant but decreasing very slowly $-\dot H/H^2 \simeq \epsilon \ll 1$. This implies that $H_*$ at the end of inflation can only be smaller than $H_{\rm max}$. Therefore, our normalization in Eq.~(\ref{eq:InitialRatio}) is actually conservative, and we rather expect typically $\Delta_* < \delta\cdot 10^{-12}$.

\subsection{Inverting the energy hierarchy}\label{subsec:hierarchyInversion}

By definition, inflation occurs when the equation of state (EoS) {\small$w$} satisfies {\small$w < -1/3$}. The EoS in the epoch after inflation must fall in the range {\small$-1/3 < w < 1$}, and even though it is common to assume that {\small$0 \leq w \leq 1/3$}, there is $a~priori$ no reason (theoretical or observational) to exclude the {\it stiff} case {\small$1/3 < w < 1$}. In fact, a post-inflationary period with a stiff EoS can be realized easily in a generic model of inflation. For example, in scalar singlet driven inflation, the slow-roll condition is achieved by simply demanding {\small$V \gg K$}, where {\small$V$} and {\small$K$} are the inflaton potential and kinetic energy densities. Inflation ends when the potential drops to {\small$V < K/2$}.  If a feature in the inflaton potential allows its value {\small $V$} to drop much below the kinetic energy {\small $K$}, the EoS can become stiff after inflation, {\small$w = (K-V)/(K+V) > 1/3$}. 

The simplest realization of this {\it Kination-dominated} (KD) regime is to assume a rapid transition of the potential from {\small$V \gg K$} during inflation to some small value {\small$V \ll K$} after inflation. The transition would actually trigger itself the end of inflation, leading to a post-inflationary EoS {\small$w \simeq 1 - \mathcal{O}(V/K)$}. In general we expect that the EoS can approach unity from below, but never achieve $w = 1$ exactly, as this would require an exactly flat direction with $V = 0$. A natural scenario where inflation is followed by a KD phase is that of Quintessential-Inflation~\cite{Peebles:1998qn}, where the inflaton potential $V(\phi)$ is engineered so that the necessary transition occurs at the end of inflation, and the potential is also adjusted to describe the observed dark energy as a quintessence field, see e.g.~\cite{Peloso:1999dm,Huey:2001ae,Majumdar:2001mm,Dimopoulos:2001ix,Wetterich:2013jsa,Wetterich:2014gaa,Hossain:2014xha,Rubio:2017gty} for different proposals. An alternative scenario where a phase of KD follows after inflation, is obtained when the inflaton potential is of the form $V \propto \phi^{2n}$ after inflation, with $n$ a large positive integer. This leads to fast inflaton oscillation after inflation, where the kinetic energy dominates  on average (per oscillation) over the potential energy. An effective (oscillation averaged) EoS $w \simeq (n-1)/(n+1)$ emerges, exhibiting stiff values for $n \geq 3$~\cite{Turner:1983he}.  Like in the previous case, we also expect that the post-inflationary EoS can approach unity only from below, as achieving $w = 1$ exactly requires $n \rightarrow \infty$. We note that, in any case, a stiff period cannot be sustained for very long in this type of constructions, as self-resonant effects lead to a fragmentation of the coherent oscillating condensate~\cite{Lozanov:2017hjm}.

For our purpose, the details of the KD phase implementation within the inflationary sector are actually irrelevant. Hence, from now on we will rather focus on the phenomenology of KD, simply assuming that there is such a phase following the end of inflation. The energy density of the inflationary sector evolves therefore, after inflation, as {\small$\rho_{\rm Inf} = 3m_p^2H_*^2(a_*/a)^{3(\bar w+1)}$}, where {\small$\bar w$} is the effective (logarithmic-averaged) value of the EoS during the stiff period, spanning from the end of inflation until the moment when the universe first becomes RD\footnote{In general the EoS is determined by the inflaton potential and it is a function of time. However we expect it to change only adiabatically during KD, and in any case we can always describe the scaling of the energy density in terms of an effective average EoS $\log(\rho_{\rm rad}/\rho_{\rm rad}^*) = -3\int {da\over a}(1+w(a)) \equiv -3(1+\bar w)\log (a/a_*)$.}. Also, we note that all particle production examples discussed in Sect.~\ref{subsec:PartProd} are excitations of relativistic $dof$, and hence their energy density scales after inflation as {\small$\rho_{\rm rad} = 3m_p^2H_*^2\Delta_*(a_*/a)^4$}. Therefore, the ratio of the energy density of the radiation species to that of the inflaton evolves as
\begin{align}\label{eq:EvolutionRatioEnergies}
\Delta(t) \equiv {\rho_{\rm rad}\over \rho_{\rm Inf}} = \Delta_*\left(a\over a_*\right)^{3\bar w-1} \,,
\end{align}
where recall that {\small$\Delta_* \ll 1$} [c.f.~Eq.~(\ref{eq:InitialRatio})] represents the initial ratio of the radiation fields' energy density to that of the inflaton. If {\small$w \leq 1/3$}, either {\small$\Delta(t)$} remains as small as {\small$\Delta_*$} ({\small$\bar w = 1/3$}), or decrease even further as {\small$\propto (a/a_*)^{-(1-3\bar w)}$} ({\small$0 \leq \bar w < 1/3$}). However, for a stiff EoS %{\small$0 < \delta w \leq 2/3$} 
{\small$1/3 < \bar w < 1$}, {\small$\Delta(t)$} grows  as {\small$\propto (a/a_*)^{(3\bar w-1)}$}. Despite starting from a very small value, {\small$\Delta(t_*) = \Delta_* \ll 1$}, for a stiff EoS there is always a time {\small$t_{\rm RD}$} for which {\small$\Delta(t \geq t_{\rm RD}) \geq 1$}. Thus, after {\small$t = t_{\rm RD}$} the energy budget of the universe becomes dominated by the radiation sector.

Setting the conformal time at the end of inflation to zero, $t_* = 0$, the scale factor evolves deep inside the Stiff epoch as
\begin{eqnarray}\label{eq:stiffscalefactor}
a(t) = a_*\left(1+ {1\over \alpha}a_*H_*t\right)^{\alpha}\,,~~~\alpha \equiv {2\over 1+3\bar w} \,,
\end{eqnarray}
with $\alpha$ ranging from 1 ($\bar w = 1/3$) to 1/2 ($\bar w = 1$). By construction {\small$1 = \Delta_* (a_{\rm RD}/a_*)^{3\bar w - 1}$}, and hence {\small$a_{\rm RD}/a_* = \Delta_*^{-{1\over 3\bar w - 1}} = \Delta_*^{-{\alpha\over 2(1-\alpha)}}$}.  Using {\small $a_{\rm RD}/a_* \simeq \alpha^{-\alpha}(a_*H_*t)^{\alpha}$} for $t \gg 1/(a_*H_*)$, we then obtain
\begin{eqnarray}\label{eq:RHtime}
t_{\rm RD} \simeq \alpha\Delta_*^{-{1\over2(1-\alpha)}}(a_*H_*)^{-1}\simeq& \alpha\,\left(10^{12}\over\delta\right)^{1\over2(1-\alpha)}\left({H_*\over H_{\rm max}}\right)^{-{1\over(1-\alpha)}}(a_*H_*)^{-1}\,.\nonumber
\end{eqnarray} 
To get an idea of the order of magnitude of this scale, let us compute its lower bound for $\delta=1$. Plugging $H_* \simeq H_{\rm max}$ and $\bar w  \simeq 1$, we obtain
\begin{align}\label{eq:RHtime}
t_{\rm RD} \geq 5\cdot 10^{11} \,(a_*H_*)^{-1}\,,
\end{align}
which represents a much longer timescale than the initial conformal Hubble time $(a_*H_*)^{-1}$. 

If the radiation species are already in thermal equilibrium when its energy dominance begins, we can compute the temperature {\small$T_{\rm RD}$} of the system at {\small$t = t_{\rm RD}$}. Using {\small$\rho_{\rm RD}(t_{\rm RD}) \equiv {\pi^2\over30}g_{_{\rm RD}}T_{\rm RD}^4$} {\small$= 3m_p^2H_*^2\Delta_*(a_*/a_{\rm RD})^4$}, we find
\begin{align}\label{eq:RHtemp}
T_{\rm RD} \simeq {4.24 \over g_{_{\rm RD}}^{{1/4}}}\cdot 10^{18}\left({\delta\over300}\right)^{1+\alpha \over 4(1-\alpha)}\left(H_*\over mp\right)^{1\over (1-\alpha)}~{\rm GeV},
\end{align}
with {\small$g_{_{\rm RD}}$} the number of thermal radiation $dof$ at {\small$t_{\rm RD}$}. The temperature {\small$T_{\rm RD}$} [Eq.~(\ref{eq:RHtemp})] can therefore be identified as the $reheating$ $temperature$, i.e.~the highest temperature reached by the thermal plasma when it first dominates the energy budget of the Universe. For {$\bar  w  \simeq 1$}, {$H_* \simeq H_*^{\rm max}$} and $\delta = 1$, we obtain a maximum reheating temperature of
\begin{eqnarray}
T_{\rm RD} \sim 10^{7} ~{\rm GeV}~~~(\bar w \simeq 1, H_* = H_{\rm max}, \delta = 1)\,,
\end{eqnarray}
So, in general, we expect $T_{\rm RD} < 10^{7} ~{\rm GeV}$.

\subsection{Minimal requisites}
\label{sec:MinimalRequisites}
\vspace*{0.1cm}

Though not always discussed in the literature, the process just described is a viable reheating mechanism only if certain non-trivial conditions are met:\\

{\it 1) Ensuring thermal equilibrium before  BBN\,-.} If the radiation fields interact among themselves, say via gauge interactions, the initially out-of-equilibrium distributions will naturally evolve towards thermal equilibrium. The equilibration time in a gauge theory can be estimated as {$\tau_{\rm th} \sim 1/(\alpha^2T_{\rm th})$}, where {\small$T_{\rm th}$} is the temperature of the system when thermal equilibrium is first established, and {\small$\alpha = g^2/4\pi$} is the relevant gauge coupling. Using {$\rho_{\rm th} = (g_{\rm th}\pi^2/30)T_{\rm th}^4 = 3m_p^2H_*^2\Delta_{*}/a_{\rm th}^{4}$}, $T_{\rm th} \sim 1/ \alpha^2\tau_{\rm th}$ and $a_{\rm th}/a_* \simeq (3H_* \tau_{\rm th})^{1/3}$ (for simplicity we consider $w \simeq 1$, $\delta = 1$ and $g_{\rm th}\pi^2/90 \sim 1$), we find $ H_*\tau_{\rm th} \sim 10/\alpha^3$. For a gauge coupling with the strength of the electroweak interactions of the Standard Model, $g^2 \simeq 0.3$, we obtain {\small$\tau_{\rm th} \sim 10^6 H_*^{-1} $}, which is much smaller than $\tau_{\text{BBN}}\sim H_{\text{BBN}}^{-1}\sim m_p/T_{\rm BBN}^{2}\sim 10^{24}\,\text{GeV}^{-1}$, unless $H_*$ is as tiny as $H_*\lesssim 10^{-18}\text{ GeV}$. While more elaborated calculations of {$\tau_{\rm th}$} can be made, see e.g.~\cite{Arnold:2002zm,Kurkela:2011ti,Kurkela:2014tea}, the precise value of {$\tau_{\rm th}$} is irrelevant for the purpose of reheating the universe into a radiation background, as long as $\tau_{\rm th} \ll \tau_{\rm BBN}$.\\

{\it 2) Ensuring small inflaton fluctuations at the end of inflation\,-}. The inflaton may undergo a non-adiabatic change in mass during the rapid transition from the inflationary (quasi-)exponential expansion to the post-inflationary power-law expansion. The inflaton fluctuations generated during this transition may be a potential threat to the gravitational reheating scenario. The question is twofold: $i)$ Is the energy fraction of the inflaton fluctuations larger than the energy fraction of the radiation produced due to their non-minimal coupling to gravity found in Eq.~(\ref{eq:InitialRatio})? $ii)$ Do the energy density of the inflaton fluctuations scale slower than the radiation component? If the answer to both questions is ``yes" then the energy of the inflaton fluctuations will forever dominate over that of the non-minimally coupled field(s), since we are assuming an isolated inflaton sector which cannot decay into other non-inflationary $dof$. This would spoil the goal of successfully reheating the Universe as we require non-inflationary $dof$ (which eventually lead to SM fields) to dominate the energy budget of the Universe at some point. Let us start with the first condition. In the fast transition limit, the energy fraction of the inflaton fluctuations immediately after the transition can be estimated as~\cite{Figueroa:2016dsc}
\begin{eqnarray}\label{eq:inflatonFraction}
\frac{\rho_{\rm Inf}^{\rm decay}}{3 m_p^2 H_*^2}\sim {m_\phi^4 \over 3 m_p^2 H_*^2} \sim \mathcal{O}(10^{-9}) \times \eta_\phi^2 \left({H_*\over H_*^{\rm max}}\right)^2
\end{eqnarray}
where $m_\phi^2 = V''(\phi)$ is the effective inflaton mass right before the transition, and $\eta_\phi\equiv m_\phi^2/(3H_*^2)<1$ is just a number evaluated right before the transition. Comparing this ratio to Eq.~(\ref{eq:InitialRatio}), we see that the inflaton fluctuations' energy is always larger than the initial energy of the radiation sector. This will be a serious problem if the second condition mentioned above is also satisfied. Since the fluctuations are only excited for $k/a \lesssim m_{\phi}$~\cite{Herranen:2015ima,Figueroa:2016dsc}, the equation of motion of the inflaton fluctuations can be written as $\ddot{\delta\phi_k} + 3(\dot a/a)\dot{\delta\phi_k} \simeq - V''(\phi_*)\delta\phi_k$, where we are neglecting the term $(k^2/a^2)\delta\phi_k$. As we can see, the equation of motion obeyed by the fluctuation $\delta \phi_k$ is the same as that obeyed by the condensate $\phi$, which means that the energy density of the velocity fluctuations decays with the same KD scaling law as that of the background field. For instance, if we take $V'' = 0$ (corresponding to $\bar w = 1$) in the $rhs$ of the above equation, the solution to the velocity fluctuation reads $\dot{\delta\phi_k} \propto 1/a^3$, and hence $\langle (\delta\dot{\phi})^2 \rangle  \sim \int dk k^2 \langle |\dot{\delta\phi_k}|^2 \rangle \propto 1/a^6$. In general we expect $\langle (\delta\dot{\phi})^2 \rangle  \sim \int dk k^2 \langle |\dot{\delta\phi_k}|^2 \rangle \propto 1/a^{3(1+\bar w)}$. This means that the kinetic energy of the inflaton fluctuations can be regarded as part of the energy budget of the inflaton condensate, and hence the fact that the fraction~Eq.~(\ref{eq:inflatonFraction}) is larger than Eq.~(\ref{eq:InitialRatio}) is actually irrelevant.\\

{\it 3) Ensuring the right cosmological perturbations\,-}. As we have seen, a sufficiently long period of KD allows the non-inflationary radiation $dof$ to eventually dominate the energy density of the Universe. When this occurs,  at {\small$t = t_{\rm RD}$}, the radiation field(s) perturbations are converted into adiabatic perturbations. This allows the possibility of generating the cosmological perturbations through the curvaton mechanism~\cite{Enqvist:2001zp,Lyth:2001nq,Moroi:2001ct,Kunimitsu:2012xx}. Suppose that the radiation field perturbations were generated through a field $\varphi$ that is excited during inflation. The curvature power spectrum generated by such field is given by $\mathcal{P}_{\zeta} \sim 0.1\delta \rho/\langle\rho\rangle$, with $\rho$ and $\delta \rho$ the mean value and the typical fluctuation amplitude of the field energy density. For a single spectator field $\varphi$ during inflation with a quadratic potential we obtain a curvature perturbation with power spectrum $P_{\zeta} \sim \mathcal{O}(1)$, so this case is strongly forbidden~\cite{Lyth:2001nq}. In the case of a quartic potential, we obtain $P_{\zeta} \sim 0.1 (\delta\varphi)^2/\langle\varphi^2\rangle \sim 0.01\sqrt{\lambda}$, where $\lambda$ is the radiation field self-coupling. Unless $\lambda$ is greatly fine-tuned to $\lambda \sim 10^{-14}$ in order to explain the observed perturbation amplitude $\mathcal{P}_{\zeta}^{\rm obs} \sim 10^{-9}$, this case is also ruled out. This conclusion certainly applies to the case of the SM Higgs, which is excluded as a curvaton~\cite{Kunimitsu:2012xx,Figueroa:2016dsc}, as the Higgs self-coupling is expected to be of the order of $\lambda \lesssim 10^{-2}$ at high energy scales (assuming stability of the SM Higgs potential). For the required small self-coupling $\lambda \sim 10^{-14}$, the number of e-folds required for the field to arrive at its equilibrium fluctuation is of the order of $\sim 1/\sqrt{\lambda} \sim 10^7$ e-folds, which makes the set-up somehow contrived, as we only need $\sim$ 50-60 e-folds of inflation to solve the horizon and flatness problems. In the case where the radiation field(s) are excited towards the end of inflation due to a non-minimal gravitational coupling, the resulting field fluctuation are rather large, with the spectrum peaked around the Hubble scale at the end of inflation, power-law and exponentially suppressed at super- and sub-horizon scales, respectively. Hence, there is no problem in this case if we generate a too large cosmological perturbations, as they only affect very small scales. Furthermore, since the unavoidable gravitational couplings between the inflaton and the radiation field(s) mean that the inflaton perturbations are preserved even after the inflaton energy density becomes negligible at {\small$t > t_{\rm RD}$}~\cite{Sloth:2005yx,Bartolo:2005jg}, the case of non-minimally coupled fields excited towards the end of inflation remains still observationally viable, provided that the inflationary sector is chosen such that it generates the observed perturbation spectrum.\\

\begin{figure}[t]
\center
\includegraphics[width=10cm]{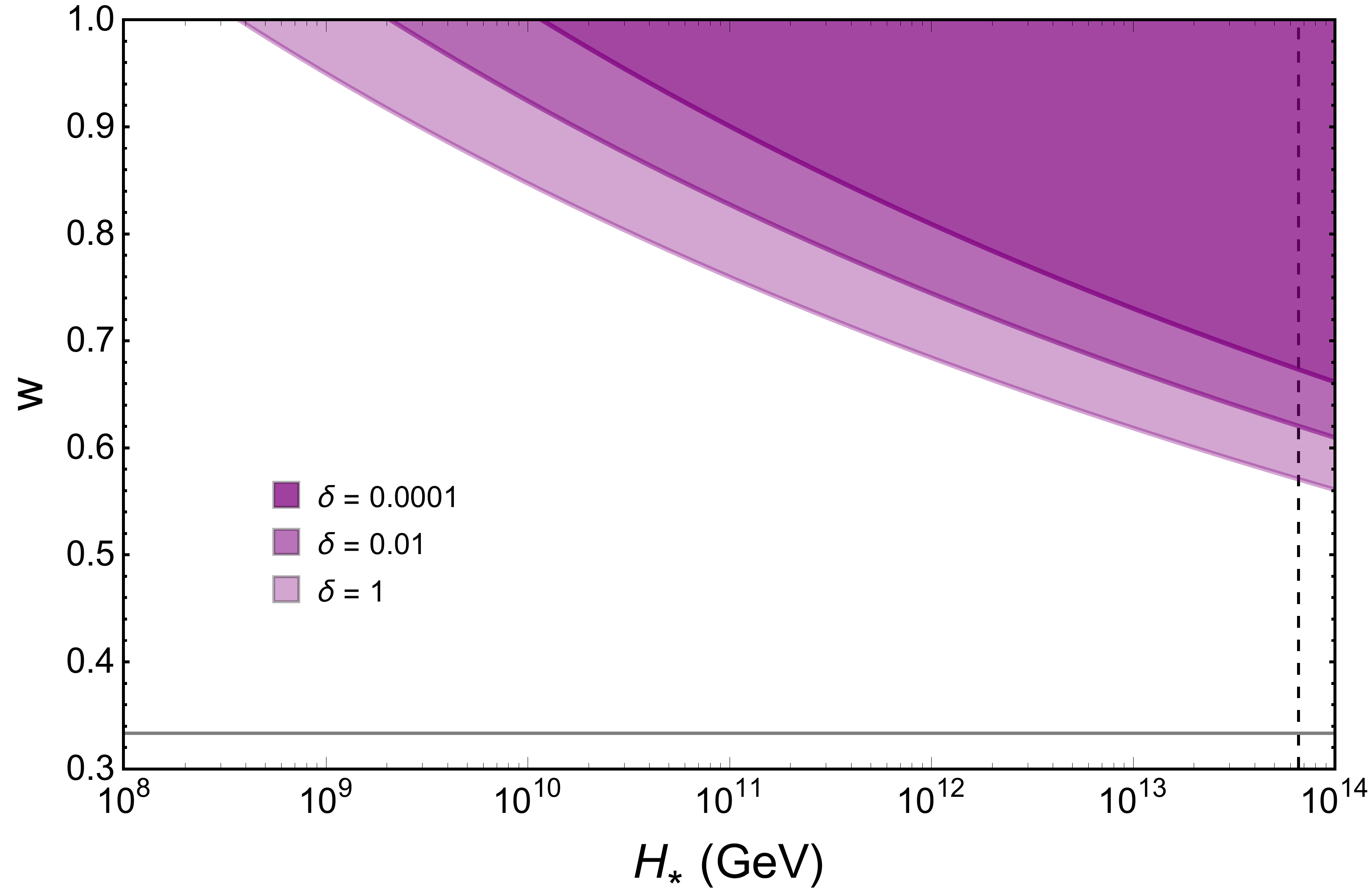}
\caption{Coloured regions correspond to $\bar w \geq {w}_{\rm RD}$. The borders of the regions depict ${w}_{\rm RD}$ as a function of $H_*$ for the three values $\delta = 0.001, 0.01, 1$, respectively coloured in dark, medium and light purple. The horizontal line near the bottom marks $w = 1/3$ while the vertical dashed line marks $H_* = H_*^{\rm max}$.}
\label{fig:1}
\end{figure}

{\it 4) Ensuring radiation domination before BBN\,-.} In order to preserve the success of BBN, we need the radiation sector to dominate before the onset of BBN\footnote{From now on we will make the loose identification of BBN with the $\sim 1$ MeV scale, even though BBN rather starts at a smaller energy scale, $\lesssim 0.1$ MeV. This is because the freeze-out of neutron/proton conversion takes place around $\sim 1$ MeV, and any change in the relativistic number of $dof$ from then on will affect the expansion rate and thus the amount of beta decay from that time till the end of BBN.}. In particular, using Eq.~(\ref{eq:RHtemp}) and imposing the condition
\begin{align}
T_{\rm RD}(\bar w,H_*;\delta) \geq T_{\rm BBN} \simeq 1\text{ MeV}\,,
\end{align}
we arrive at the condition
\begin{align}
{4.24 \over g_{_{\rm RD}}^{{1/4}}}\cdot 10^{21}\left({\delta\over300}\right)^{1+\alpha \over 4(1-\alpha)}\left(H_*\over mp\right)^{1\over (1-\alpha)} \geq 1\,,
\end{align}
which implies that
\begin{equation}
\bar w \geq {w}_{\rm RD}(H_*;\delta)\,,
\end{equation}
with ${w}_{\rm RD}(H_*;\delta)$ a function of $H_*$ and $\delta$, see Fig.~\ref{fig:1}. We obtain this condition by considering the number of relativistic $dof$ at the onset of BBN as $g_{\rm RD} = g_{\rm BBN} \simeq 2 + (7/8)\cdot2\cdot3\cdot(4/11)^{4/3} \simeq 3.36$. For {\small$H_* = H_{\rm max}$} and {\small$\delta = 1$}, we have
\begin{align}
\bar w \geq {w}_{\rm RD} \simeq 0.57\,,~~~ (\delta = 1, H_* = H_{\rm max})\,.
\end{align}
For a given value of $\delta$, the value of ${w}_{\rm RD}$ increases as we decrease the inflationary Hubble rate $H_* < H_{\rm max}$, since the initial energy ratio $1/\Delta_* \gg 1$ to overcome becomes larger. When we decrease $\delta$, ${w}_{\rm RD}$ increases again, as this also increases $1/\Delta_* \gg 1$.  If, for a fixed value of $\delta$, we continually decrease $H_*$, we will eventually hit ${w}_{\rm RD} = 1$, so that the corresponding inflationary Hubble rate  $H_{\rm min}$ at that point is the minimum value we can tolerate for the given $\delta$. For $\delta = 1$ we find $H_{\rm min} \simeq 4\cdot10^{8}$ GeV, and for smaller values of $\delta$, $H_{\rm min}$ increases. If $H_* < H_{\rm min}$ we can never enter the RD epoch before the onset of BBN. See Fig.~\ref{fig:1}, where we plot ${w}_{\rm RD}$ as a function of $H_*$ for the three values $\delta = 10^{-4}, 10^{-2}, 1$. \\

In summary, several properties must be met for gravitational reheating to be a cosmologically viable scenario. While how fast the radiation fields thermalize [point $1)$ above] is a model dependent question, the equality of the redshift rate of the energy density of the inflaton velocity fluctuations and that of the background make the former harmless [point $2)$]. The requirement to produce the right cosmological perturbations [point $3)$] imposes strict conditions on the properties of the radiation fields that will eventually dominate the energy budget of the Universe. Finally, ensuring that RD starts before BBN [point $4)$] imposes a severe restriction on the inflationary sector, demanding $w_{\rm RD} < \bar w < 1$, so that the EoS after inflation must  be rather closer to $1$ than to $1/3$.

As a final remark, let us note that $\delta \gg 1$ is only possible if we assume the existence of a rather large number of species forming the radiation sector. However, as explained in point $3)$, the characteristic coupling of each species need to be tuned to very specific values in order not to spoil currently observed properties of the cosmological perturbations. If the fields are excited during inflation out of quantum vacuum fluctuations, free light massive $dof$ are ruled out, whereas self-interacting fields require a fine-tuning of their self-coupling; if the fields are excited towards the end of inflation through a non-minimal gravitational coupling, the inflaton potential must provide the origin of the cosmological perturbation, while the non-minimal coupling must be tuned to $|1-6\xi| \sim 1$. There seems not to be any particular reason to expect that hundreds of field species should exist during inflation with their coupling appropriately tuned to avoid spoiling the cosmological perturbation, with the only purpose that the amount of energy stored in the radiation component at the end of inflation can be enhanced (linearly in the number of fields).  Invoking such a seemingly carefully-engineered case does not seem very appealing, reinforcing that $\delta\leq 1$ is the canonical range, whereas $\delta \gg 1$ requires rather unmotivated $ad-hoc$ constructions.

\section{The problem with gravitational reheating}

We will assume from now on that cosmologically viable scenarios of gravitational reheating exist, satisfying points $1)-4)$ from Sect.~\ref{sec:MinimalRequisites},  without entering into the details of their particle physics implementation. A gravitational reheating scenario can then be characterized by three parameters: $\delta$, controlling the initial ratio of the energy density of radiation to that of the inflaton; $H_*$, corresponding to the Hubble rate at the end of inflation; and $\bar w$, representing the averaged EoS in the stiff epoch following inflation. The inflationary Hubble rate is constrained as $H_* \leq H_{\rm max}$, whereas the typical value of the EoS lies in the range $0.57 \lesssim w_{\rm RD}  \leq \bar w < 1$. 

In this section, we will show that any gravitational reheating scenario with $\delta$ lying within its generic range $\delta \leq 1$ is actually inconsistent. The inconsistency arises mainly due to the need to preserve the success of BBN, particularly that the total energy density of extra relativistic species (which are not part of the SM) present during BBN must not contribute significantly to the expansion rate of the Universe, or otherwise the resulting light element abundances would be spoiled. Since GWs are one type of such extra relativistic species, the energy density of the stochastic background of GWs should be smaller than the upper bound~\cite{Caprini:2018mtu}
\begin{equation}\label{eq:BBNbound}
\left(\frac{h^2\rho_{\rm GW}}{\rho_c}\right)_0 = \int \frac{df}{f} \, h^2 \, \Omega_{\rm GW}(f) \leq 5.6 \times 10^{-6} \, 
\Delta N_{\nu} \,,
\end{equation}
where $\Delta N_{\nu}$ parametrizes the extra amount of radiation from beyond the SM $dof$\footnote{The contribution from extra radiation during BBN or other stages of the evolution of the universe is typically parametrized in terms of an effective deviation $\Delta N_{\nu}$ from the number of SM neutrino species $N_{\nu}=3$. This is only a parametrization. The extra radiation does not need to be neutrinos and can be either bosonic or fermionic.}~\cite{Mangano:2005cc}. As we will see soon, a post-inflationary stiff epoch breaks the (quasi-)scale invariance of the GW background from inflation, inducing a positive tilt in the high-frequency part of the spectrum corresponding to the modes that crossed the horizon during the stiff epoch. The stiffer the EoS $\bar w$, the steeper the tilt. %, and hence the larger the total GW energy density. Hence, for a given scenario characterized by some values of $\delta$ and $H_*$, there is always an upper bound for the stiff EoS $w_{\rm BBN}(\delta, H_*)$ above which Eq.~(\ref{eq:BBNbound}) is violated. 
In what follows, we will show that there is no choice of parameters $\lbrace \delta, H_*, \bar w \rbrace$ within the canonical domains $\delta \leq 1, H_* \leq H_{\rm max}, w_{\rm RD} \leq \bar w < 1$, for which Eq.~(\ref{eq:BBNbound}) is not violated. %$\bar w < w_{\rm BBN}(\delta,H_*)$. 

\subsection{Distortion of the inflationary GW background due to the stiff period}\vspace*{0.1cm}

Let us first of all recall that GWs are transverse ($\partial_ih_{ij} = 0$) and traceless ($h_{ii} = 0$) metric perturbations $ds^2 = a^2(t)(-dt^2+(\delta_{ij}+h_{ij})dx^idx^j)$. Their energy density spectrum (at sub-horizon scales) is defined as~\cite{Caprini:2018mtu}
\begin{align}
\Omega_{\text{GW}}(t, k) \equiv\dfrac{1}{\rho_{\text{crit}}}\dfrac{d\rho_{\text{GW}}(t,k)}{d\ln k} =\frac{k^2}{12a^2(t)H^2(t)}\Delta_h^2(t,k)\,,  \label{eq:GWenergySpectrum}
\end{align}
where $\Delta_h^2(t,k)$ is the tensor power spectrum at arbitrary times, defined as
\begin{align}
\left<h_{ij}(t,\mathbf{x})h^{ij}(t,\mathbf{x})\right>&\equiv\int \frac{dk}{k} \Delta_h^2(t,k)\,,
\end{align}
with $\left<...\right>$ denoting an average over a statistical ensemble. In our case, we can factorize the tensor power spectrum as
\begin{align}
\Delta_h^2(t,k)\equiv T_h(t,k) \Delta_{h,\text{inf}}^2(k)\,, \label{transferfunc}
\end{align}
with  $T_h(t,k)$ a transfer function and $ \Delta_{h,\text{inf}}^2(k)$ the primordial tensor spectrum from inflation
\begin{align}
\Delta_{h,\text{inf}}^2(k) \simeq {2\over \pi^2}\left({H_{\rm inf}\over m_p}\right)^2\left({k\over k_{p}}\right)^{n_t}\,,
\label{eq:InfSpectrum}
\end{align}
with $n_t$ a spectral tilt, $k_{p}$ a pivot scale of the order the Hubble rate at the time of CMB decoupling, and $H_{\rm inf}$ the Hubble rate when the mode $k_p$ exited the Hubble radius during inflation. The upper bound $H_{\rm max}$ introduced in Sect.~\ref{sec:GravRHidea} actually applies only to the Hubble rate $H_{\rm inf}$, i.e. $H_{\rm inf} \leq H_{\rm max}$, rather than to $H_*$, as we already explained below Eq.~(\ref{eq:Hmax}). %, particularly for $(a_p/a_0)k_{p} \simeq 0.002~{\rm Mpc}^{-1}$. 
The spectrum is expected to be only slightly red-tilted in slow-roll inflation, with the spectral index being 'slow-roll suppressed' as
\begin{align}\label{eq:consistencyCondition}
n_t \simeq -2\epsilon \simeq -{r_{0.002}\over8} \,,
\end{align}
where $r_{0.002} \equiv \Delta_h^2/\Delta_{\mathcal R}^2$ is the tensor-to-scalar ratio evaluated at the scale $k_p$, constrained by the most recent analysis by the Planck collaboration as $r_{0.002} \leq 0.064$~\cite{Akrami:2018odb}. It is actually this bound that determines the upper bound $H_{\rm max}$ on the inflationary Hubble scale reported in Eq.~(\ref{eq:Hmax}). Given this bound, we see that the inflationary spectrum has a very small red-titled spectral index $-n_t \leq 0.008 \ll 1$, which makes the spectrum very close to exact scale-invariance, at least around the CMB scales\footnote{For instance, with no running of the spectral index, the amplitude of the tensor spectrum would fall only by a factor $\sim (10^{25})^{-0.008} \sim 0.6$ during the $\ln(e^{60}) \sim $ 26 orders of magnitude separating the CMB scales and the scale that matches the Hubble radius at the end of inflation.}. From now on, for simplicity of our discussion, we will consider an exact scale-invariant inflationary spectrum, as this gives an excellent approximation. We will comment on deviations from this assumption in Sect.~\ref{sec:Discussion}. 

Let us assume for a moment that, immediately after inflation, the Universe became RD with equation of state $w = 1/3$. The resulting present-day GW energy density spectrum would then be scale-invariant for the frequency range corresponding to the modes crossing the Hubble radius during RD. Using Eqs.~(\ref{eq:GWenergySpectrum}), \eqref{transferfunc}, and (\ref{eq:InfSpectrum}) (with $n_t = 0$), we obtain
\begin{eqnarray}\label{eq:InfGWtodayRD}
\Omega_{\rm GW}^{(0)}{\Big |}_{\rm RD} \simeq { \Omega_{\rm rad}^{(0)}\over12\pi^2}\left(g_{s,0}\over g_{s,k}\right)^{4\over3}\left(g_{*,k}\over g_{*,0}\right)\left(H_{\rm inf}(k)\over m_p\right)^2\nonumber\simeq 1\cdot10^{-16}\left(H_{*}\over H_{\rm max}\right)^2\,,
\end{eqnarray}
where in the first equality we have introduced the RD transfer function~\cite{Boyle:2005se,Watanabe:2006qe} 
\begin{align}
T(k) \simeq {1\over2}\left({a_k\over a_0}\right)^2 \hspace*{-0.2cm}\simeq {1\over2}\left(g_{s,0}\over g_{s,k}\right)^{4\over3}\left(g_{*,k}\over g_{*,0}\right)\Omega_{\rm rad}^{(0)}\left({a_0H_0\over a_kH_k}\right)^2\,,
\end{align}
where the prefactor $1/2$ is due to oscillation-averaging the tensor mode functions. In this way, we obtain ${k^2\over a_0^2H_0^2} T(k) \simeq {1\over2}\Omega_{\rm rad}^{(0)}\left(g_{*,k}\over g_{*,0}\right)\left(g_{s,0}\over g_{s,k}\right)^{4/3}$, where we have used $k = a_kH_k$. In the second line of Eq.~(\ref{eq:InfGWtodayRD}) we have used $\Omega_{\rm rad}^{(0)} \simeq 9\cdot 10^{-5}$, $H_{\rm inf} \simeq H_*$, $g_{s,0} \simeq 3.91$, $g_{*,0} = 3.36$ and $g_{s,k} \simeq g_{*,k} \simeq 100$.

Eq.~(\ref{eq:InfGWtodayRD}) describes the amplitude of the scale-invariant $plateau$ of the inflationary GW energy density spectrum today, corresponding to the modes that crossed the horizon during RD. If prior to RD, there is a stiff period with EoS $\bar w > 1/3$, as we have discussed before for the implementation of gravitational reheating, the resulting GW energy density spectrum will not remain scale-invariant. The spectrum today will actually consist of two parts: a blue-tilted high-frequency branch, corresponding to the modes that crossed the horizon during the stiff epoch \cite{Giovannini:1998bp}, and a scale-invariant branch corresponding to the modes that crossed the horizon during RD\footnote{There is yet another part of the spectrum, corresponding to modes that crossed the Hubble radius after matter-radiation equality, which behaves as $\Omega^o_{\rm GW} \propto 1/k^2$. Nevertheless, this corresponds to very small frequencies today $f \lesssim 10^{-16}$ Hz. We will not be concerned with such low frequency end of the spectrum, as it only affects the CMB and it cannot be probed by direct-detection experiments.}. The rationale is very simple: during the post-inflationary evolution of the universe, the tensor modes excited during inflation [with spectrum Eq.~(\ref{eq:InfSpectrum})] are frozen while they remain super-horizon. As the universe expands, successive modes become sub-horizon, and subsequently behave as GWs with amplitudes scaling as $h_{ij} \propto 1/a$. The energy density of the sub-horizon modes then scales as ${d\rho_{\rm GW}\over d\log k} \propto 1/a^4$, i.e. the scaling law of relativistic $dof$. Hence, the ratio of the GW energy density spectrum to the inflationary background with EoS $\bar w$,  for a fixed mode $k$, scales as ${1\over \rho_{\rm Inf}}{d\rho_{\rm GW}\over d\log k} \propto a^{3\bar w -1}$, which is a growing function for a stiff EoS $\bar w > 1/3$. As successive modes cross the horizon during the stiff epoch, the spectrum becomes ${1\over \rho_{\rm Inf}}{d\rho_{\rm GW}\over d\log k} \propto k^2a^{2}(t_k) \propto k^2 t_k^{2\alpha} \propto k^{2(1-\alpha)}$, where we have used that at horizon crossing $k = a_kH_k = {\alpha t_k^{-1}}$, with $\alpha$ an $\mathcal{O}(1)$ constant defined below in Eq.~(\ref{eq:alphaDef}) [c.f.~Eq.~(\ref{eq:stiffscalefactor})]. 

A detailed computation (see Appendix \ref{appendix:GWspectrumderivation}) shows that the GW energy density spectrum today is proportional to  $\Omega_ {\rm GW}^{(0)}{\Big |}_{\rm RD}$ [c.f.~Eq.~(\ref {eq:InfGWtodayRD})], modulated by a window function $\mathcal{W} (f;\bar w)$  due to the stiff epoch,
\begin {align}\label{eq:FinalSpectrum}
\Omega_ {\rm GW}^{(0)}(f) \simeq  \Omega_ {\rm GW}^{(0)}{\Big |}_{\rm RD}\times \mathcal{W}(f;\bar w)\,.
\end {align}
Defining the frequency of the mode crossing the horizon at the onset of RD by the condition $a_{\rm RD}H_{\rm RD} = k_{\rm RD} \equiv 2\pi a_o f_{\rm RD}$  (we provide explicit expressions in the next subsection), one obtains that $\mathcal{W}(f ; \bar w)$ varies smoothly in the frequency range around $f = f_{\rm RD}$, approaching asymptotically unity at small frequencies $f \ll f_{\rm RD}$, 
\begin{align}
\mathcal{W}(f \ll f_{\rm RD} ;\bar w) \longrightarrow  1 \,,
\end{align}
and behaving at large frequencies $f \gg f_{\rm RD}$ like
\begin{eqnarray}\label{eq:WlargeFreq}
&& \mathcal{W}(f \gg f_{\rm RD}; \bar w) \longrightarrow \mathcal{A}_{\alpha}\left(f\over f_ {\rm RD} \right)^{2(1-\alpha)}\,,
\end{eqnarray}
with
\begin{eqnarray}
&& \mathcal{A}_{\alpha} \equiv {\Gamma^2(\alpha+{1\over2})\over2^{(1-\alpha)}\alpha^{2\alpha}\Gamma^2({3\over2})} \,,~~~~ \alpha\equiv {2\over1 + 3 \bar w}\,.\label{eq:alphaDef}
\end{eqnarray}
We note that the coefficient in Eq.~(\ref{eq:WlargeFreq}) ranges as $1  <  \mathcal{A}_\alpha < 2^{5/2}/\pi \simeq 1.8$ for $1/3 < {\bar w} < 1$. Therefore, the GW energy density spectrum corresponding to modes crossing the horizon during the stiff epoch grows with frequency. For example, for $\bar w \simeq 1$ we obtain $\alpha \simeq 1/2$, and hence the high-frequency branch $f \gg f_{\rm RD}$ of the spectrum  grows linearly\footnote{Strictly speaking, if the EoS is $w = 1$ identically, there is a logarithmic correction in the GW spectrum~\cite{Giovannini:1998bp}, but as we explained before, we never really take $w = 1$ exactly, and rather consider values close to but somewhat smaller than unity.} with the frequency $\Omega_{\rm GW}^{(0)}  \propto (f/f_{\rm RD})$. Depending on the initial $H_*$, $\bar w$, and $f_{\rm RD}$, the growing spectrum may grow by a very large factor at large frequencies. 

In the next section we will confront the amplitude of the blue-tilted GW spectrum just found against upper bounds on the presence of stochastic GW backgrounds. Before we proceed, we introduce first some relations that will prove useful for our analysis. In particular, relevant ratios of scale factors are obtained as\footnote{We assume for simplicity that the number of $dof$ in the radiation sector does not change between the thermalization time $t_{\rm th}$ and the onset of RD at $t_{\rm RD} > t_{\rm th}$. Otherwise, Eq.~(\ref{eq:a*2aRD}) should pick up in the $rhs$ a factor $(g_{s,{\rm RD}}/g_{s,{\rm th}})^{1/3}(g_{*,{\rm th}}/g_{*,{\rm RD}})^{1/4}$.}
\begin{align}
\label{eq:a*2aRD}
{a_{*}\over a_{\rm RD}} &= \Delta_*^{\alpha\over2(1-\alpha)} = \left({H_{\rm RD}\over \sqrt{2}H_*}\right)^{\alpha\over(1+\alpha)}\\
\label{eq:aRd2aBBN}
{a_{\rm RD}\over a_{\rm BBN}} &= \left(g_{s,{\rm BBN}}\over g_{s,{\rm RD}}\right)^{1\over3}\left(g_{*,{\rm RD}}\over g_{*,{\rm BBN}}\right)^{1\over4}\left({\sqrt{2}H_{\rm BBN}\over H_{\rm RD}}\right)^{1\over2}\\
\label{eq:aRd2a0}
{a_{\rm RD}\over a_0} &= \left(g_{s,0}\over g_{s,{\rm RD}}\right)^{1\over3}\left(g_{*,{\rm RD}}\over g_{*,0}\right)^{1\over4}\left({\sqrt{2}H_0\over H_{\rm RD}}\right)^{1\over2}\Omega_{\rm rad}^{(0)^{1\over 4}}\,,
\end{align}
where the factors of $\sqrt{2}$ reflect the fact that we define the onset of RD by the condition $\Delta(t_{\rm RD}) \equiv 1$, which implies that $\rho_{\rm inf}(t_{\rm RD}) = \rho_{\rm rad}(t_{\rm RD})$, and hence $ H_{\rm RD}^2  \equiv H^2(t_{\rm RD}) \equiv 2\times{\rho_{\rm rad}(t_{\rm RD})\over 3m_p^2}$. The Hubble rate at $t_{\rm RD}$ and at the onset of BBN are then given by
\begin{align}
\label{eq:HRD}
H_{\rm RD}^2 &= 2\times{\pi^2\over 90}g_{*,{\rm RD}}{T_{\rm RD}^4\over m_p^2} = {2\rho_*\Delta_*\over3m_p^2}\left(a_*\over a_{\rm RD}\right)^4 \\
&= 2H_*^2\Delta_*^{1+\alpha \over 1-\alpha}= 2H_*^2\left(\delta\over300\right)^{1+\alpha \over 1-\alpha}\left(H_*\over m_p\right)^{2(1+\alpha) \over 1-\alpha}\hspace*{-6mm},\\
\label{eq:Hbbn}
H_{\rm BBN}^2 &= {\pi^2\over90}g_{*,{\rm BBN}}{T_{\rm BBN}^4\over m_p^2} \simeq (2.5\cdot 10^{-25}~{\rm GeV})^2\hspace*{-2mm}\,,   
\end{align}
where in the last equality we have used $T_{\rm BBN} \simeq 10^{-3}$ GeV at the onset of BBN, and the effective thermal relativistic $dof$ (accounting for photons and neutrinos of the SM) is $g_{\rm BBN} \simeq 3.36$. We note that we are assuming that $T_{\rm RD} \gg T_{\rm BBN}$, otherwise if the universe reheats just about before the onset of BBN, at a temperature $T_{\rm RD} \gtrsim T_{\rm BBN}$, then Eq.~(\ref{eq:Hbbn}) should pick up a factor $2$ on the $rhs$, which would cancel out the factor $\sqrt{2}$ in Eq.~(\ref{eq:aRd2aBBN}).

\subsection{Incompatibility with upper bounds on stochastic gravitational wave background}\vspace*{0.1cm}

The first problem with gravitational reheating that we immediately encounter, is that the energy density associated to the tensor modes that first crossed the Hubble radius after the onset of the stiff era is larger than the energy density of the radiation component excited during or towards the end of inflation. This can be easily seen by computing the GW spectral energy density associated to the highest frequency modes, $k\sim k_* = a_*H_*$, as their contributions dominate the integration over all the sub-horizon modes contributing to the total GW energy density. In particular, some time after such modes have entered the horizon, say once they can be considered as propagating (hence oscillating) GWs, the ratio of their logarithmic energy density spectrum to the critical energy density, can be found as
\begin{align}\label{eq:DeltaGW}
\Omega_{\rm GW}(t,k_*) \equiv {1\over \rho_c(t)}{d\rho_{\rm GW}\over d\log k}(t,k_*)  
= {k_*^2\Delta_h^2(t,k_*)\over 12 a^2(t)H^2(t)} \simeq {\mathcal{A}_\alpha\over 12\pi^2}\left(H_*\over m_p\right)^2\left(a(t)\over a_*\right)^{3\bar w -1}%\times f(\alpha)\,,\nonumber
\end{align}
where we have used the expression of $\Delta_h^2(t,k_*)$ from Appendix~\ref{appendix:GWspectrumderivation}, and $H^2(t) = H_*^2(a_*/a(t))^{2(1+\alpha)/\alpha}$. Comparing Eq.~(\ref{eq:DeltaGW}) with Eq.~(\ref{eq:EvolutionRatioEnergies}), we realize that they have not only the same time-dependence $(a/a_*)^{3\bar w -1} = (a/a_*)^{2(1-\alpha)/\alpha}$ (as expected for the ratio of energy densities of relativistic species to that of a stiff background), but also the same Hubble scale dependence $\propto (H_*/m_p)^2$. Hence, the ratio between Eq.~(\ref{eq:DeltaGW}) and Eq.~(\ref{eq:EvolutionRatioEnergies}) is time- and scale-independent,
\begin{equation}\label{eq:deltaGW}
{\Omega_{\rm GW}(t,k_*)\over \Delta(t)} \simeq {300\cdot \mathcal{A}_\alpha \over 12\pi^2\delta}\,.
\end{equation}
We observe that for the generic values $\delta \leq 1$,  ${\Omega_{\rm GW}(t,k_*)\over \Delta(t)} > 1$, i.e.~the energy density stored in the high frequency modes of the GW spectrum is larger than the energy density of the initial radiation component, c.f.~Eq.~(\ref{eq:EvolutionRatioEnergies}). Since the energy density of GWs scales as radiation, it would remain to be the dominant radiation component today, in clear conflict with the idea that the radiation sector need to be become eventually dominated by SM light species.

In fact, technically speaking, this already invalidates the analysis of gravitational reheating done before, because the universe would reheat at an earlier time than inferred from Eq.~(\ref{eq:EnergyParam}), and the universe would enter into RD with its energy budget dominated by GWs. Therefore, Eq.~(\ref{eq:deltaGW}) simply indicates that, for consistency, gravitational reheating could only be a viable scenario if $\delta$ is larger than a critical value, which in this case is given by
\begin{equation}
\delta \gg \delta_c =\frac{25}{\pi^2 }\mathcal{A}_\alpha \sim 5\,,
\end{equation}
which lies outside the generic range $\delta \leq 1$.

This leads into a similar but yet more severe problem with gravitational reheating: its incompatibility with BBN. The GW bound in Eq.~(\ref{eq:BBNbound}) applies on the total integrated GW energy density. However, except for GW spectra with a very narrow peak of width $\Delta f \ll f$, the above bound can be interpreted as a direct bound on the amplitude of a GW spectrum $h^2\Omega_ {\rm GW}^{(0)}(f) \lesssim 5.6 \times 10^{-6} \Delta N_\nu$ over a wide frequency range. This, of course, only applies to GWs with wavelengths well inside the horizon before the onset of BBN\footnote{Tensor modes with super-horizon wavelengths do not act as relativistic $dof$ since they do not propagate as waves, so they do not affect the expansion rate of the Universe at the onset of BBN.}. Hence, the bound in Eq.~\eqref{eq:BBNbound} applies only to tensor modes that crossed the horizon before the start of BBN, which correspond to present-day frequencies $f \geq f_{\rm BBN}$, with $f_{\rm BBN}$ the frequency corresponding to the (redshifted) horizon scale at the onset of BBN,
\begin{align}\label{eq:fBBN}
f_{\rm BBN} &\equiv {1\over2\pi}{a_{\rm BBN}\over a_{\rm RD}}{a_{\rm RD}\over a_0}{H_{\rm BBN}\over{\rm GeV}}\times 1.52\cdot 10^{24}\,{\rm Hz}\nn
&\simeq 2.1\cdot 10^{11}\left(g_{s,0}\over g_{s,{\rm RD}}\right)^{1\over3}\left(g_{*,{\rm RD}}\over g_{*,0}\right)^{1\over4}{\Omega_{\rm rad}^{(0)^{1\over4}}}\sqrt{{H_0\over {\rm Hz}}{H_{\rm BBN}\over {\rm GeV}}}\,{\rm Hz} \nn
&\simeq 1.6 \times 10^{-11}~{\rm Hz}\,.   
\end{align}
%where Eqs.~\eqref{eq:aRd2aBBN} and \eqref{eq:aRd2a0} were used to get to the second line. 
In going to the last line, we have used 
$H_0 \simeq 67.8\times 3.24\cdot 10^{-20}$ Hz,  $\Omega_{\rm rad}^{(0)} \simeq 9\cdot 10^{-5}$, and $g_{s,0} \simeq 3.91, g_{*,0} = g_{*,{\rm BBN}} \simeq 3.36$. In order to respect the BBN bound from Eq.~(\ref{eq:BBNbound}), we need to demand that the high frequency part of Eq.~(\ref{eq:FinalSpectrum}) satisfies 
\begin{align}\label{eq:BBNlocal}
h^2\Omega_{\rm GW}(f \gg f_{\rm RD}) 
< 1.12 \times 10^{-6}\,,
\end{align}
where we have used the most recent constraint on the number of extra relativistic species $\Delta N_\nu \lesssim 0.2$ at $95\%~C.L.$~\cite{Cyburt:2015mya}.  The condition given by Eq.~(\ref{eq:BBNlocal}) must be satisfied for all frequencies $f_{\rm RD} \ll f \leq f_*$,  where 
\begin{align}
f_{\rm RD} &={1\over2\pi}{a_{\rm RD}\over a_0}{H_{\rm RD}\over{\rm GeV}}\times1.52\cdot 10^{24}~{\rm Hz}\nn
&\simeq 2.3\cdot 10^{10}\left(g_{s,0}\over g_{s,{\rm RD}}\right)^{1\over3}\left(g_{*,{\rm RD}}\over g_{*,0}\right)^{1\over4} \sqrt{{H_o\over {\rm Hz}}{H_{\rm RD}\over {\rm GeV}}}\,{\rm Hz} \nn
&\simeq 6.3\cdot 10^{10}\left(g_{s,0}\over g_{s,{\rm RD}}\right)^{1\over3}\left(g_{*,{\rm RD}}\over g_{*,0}\right)^{1\over4} \Delta_*^{1+\alpha \over 4(1-\alpha)}\sqrt{{H_*\over m_p}}\,{\rm Hz}\nn
&\simeq 5.7\cdot 10^{10}\left(\delta\over300\right)^{1+\alpha \over 4(1-\alpha)}\left({H_*\over m_p}\right)^{1\over 1-\alpha}\,{\rm Hz}\,,
\label{eq:fRD}
\end{align}
is the frequency today corresponding to the redshifted horizon scale at the onset of RD. As expected, $f_{\rm RD}$ depends on $H_*,\bar w$ and $\delta$. The high-frequency end of the spectrum, corresponding to the mode  crossing the horizon at the onset of the stiff epoch, $a_*H_* = k_*$, can also be obtained as
\begin{align}\label{eq:f*}
f_*  &= {1\over2\pi}{a_*\over a_{\rm RD}}{a_{\rm RD}\over a_0}{H_*\over{\rm GeV}}\times1.52\cdot 10^{24}~{\rm Hz}\nn
&\simeq 4.4\cdot 10^{10}\left(g_{s,0}\over g_{s,{\rm RD}}\right)^{1\over3}\left(g_{*,{\rm RD}}\over g_{*,0}\right)^{1\over4}\left({H_{*}\over m_p}\right)^{1\over2}\Delta_*^{-{1\over4}}\,{\rm Hz} \nn
&\simeq 1.7 \cdot 10^{11}\times \delta^{-{1/4}}~{\rm Hz}\,.
\end{align}
Remarkably, $f_*$ is independent of $\bar w$ and $H_*$,  and is only (mildly) modulated by $\delta$.

As $f_{\rm RD}$ is completely determined once we specify the values of $\lbrace \delta, H_*, \bar w\rbrace$, it is convenient for the following analysis to consider alternatively $\lbrace H_*, \bar w, f_{\rm RD} \rbrace$ as the set of independent variables. For specific values of the latter, $\delta$ is univocally determined. Now, we would like to determine the regions in the parameter space $\lbrace H_*, \bar w, f_{\rm RD} \rbrace$ where the BBN condition~(\ref{eq:BBNlocal}) is satisfied. In Fig.~\ref{fig:2}, we show the 2-dimensional parameter regions satisfying condition~(\ref{eq:BBNlocal}), while varying the third variable at discrete values. We show the regions in the $\lbrace \bar w, f_{\rm RD} \rbrace$ plane for different values of $H_*$ (top panel) and the regions in the $\lbrace \bar w, H_* \rbrace$ plane for different values of $ f_{\rm RD}$ (bottom panel). From the two panels we learn that for each pair of $\lbrace f_{\rm RD}, H_* \rbrace$ values, the EoS can only be smaller than an upper bound $w_{\rm BBN}(f_{\rm RD},H_*)$, which satisfies the following properties: $w_{\rm BBN}(f_{\rm RD}^{(1)},H_*) < w_{\rm BBN}(f_{\rm RD}^{(2)},H_*)$ for $f_{\rm RD}^{(1)} < f_{\rm RD}^{(2)}$ and $w_{\rm BBN}(f_{\rm RD},H_*^{(1)}) > w_{\rm BBN}(f_{\rm RD},H_*^{(2)})$ for $H_*^{(1)} < H_*^{(2)}$. The upper bound $\bar w \leq w_{\rm BBN}(H_*,f_{\rm RD})$ can be obtained simply from the condition
\begin{align}\label{eq:wBBN}
h^2\Omega_ {\rm GW}^{(0)}{\Big |}_{\rm RD}\mathcal{A}_\alpha\left(f_*\over f_{\rm RD}\right)^{2(1-\alpha)} \leq 1.12 \times 10^{-6}\,.
\end{align}
This simple inequality determines the top boundaries of the regions depicted in Fig.~\ref{fig:2}., so the coloured areas below such boundaries represent the parameter space compatible with  the BBN bound~(\ref{eq:BBNlocal}).

From Eqs.~(\ref{eq:a*2aRD}) and ({\ref{eq:HRD}) we obtain the ratio
\begin{align}\label{eq:f*2fRDratio}
{f_*\over f_{\rm RD}} = {a_*H_*\over a_{\rm RD}H_{\rm RD}} = {1\over\sqrt{2}}\Delta_*^{-{1\over 2(1-\alpha)}} \,,
\end{align}
which we can insert in Eq.~(\ref{eq:wBBN}), yielding
\begin{align}\label{eq:deltaCondition}
\left(g_{s,0}\over g_{s,k}\right)^{4\over3}\left(g_{*,k}\over g_{*,0}\right){h^2\Omega_{\rm rad}^{(0)}\over12\pi^2}\left(H_*\over m_p\right)^2{2^{\alpha}\mathcal{A}_\alpha\over2\Delta_*} \leq 1.12 \times 10^{-6},
\end{align}
where we have used Eq.~(\ref{eq:InfGWtodayRD}). As according to Eq.~\eqref{eq:InitialRatio}, $\Delta_* =  (\delta/300)(H_*/m_p)^2$, we discover that in the above condition the dependence on $H_*$ drops out.  We arrive therefore at the following condition: in order to respect the BBN constraint~(\ref{eq:BBNlocal}), $\delta$ must be larger than a critical value
\begin{align}\label{eq:deltaCritical}
\delta \geq \delta_c \equiv 18.8\cdot2^{\alpha}\mathcal{A}_\alpha \gg 1
\,,
\end{align}
where we have used $h^2\Omega_{\rm rad}^{(0)} \simeq 4.2\cdot10^{-5}$, and $g_{*,0} = 3.36$, $g_{s,0} \simeq 3.91$, $g_{*,k} \simeq g_{s,k} \simeq 100$. The critical value $\delta_c$ is a monotonous function of $\alpha$. For an EoS just barely stiff, say $\bar w \gtrsim 1/3$, we obtain $\delta_c \simeq 37.5$, whereas for $\bar w \simeq 1$, we obtain $\delta_c \simeq 48$. In other words, we always need $\delta \gtrsim 40-50$. This reinforces the conclusion that GW reheating scenarios with \text{generic} values $\delta \leq 1$ are ruled out. 

\begin{figure}
	\center
	\includegraphics[width=10cm]{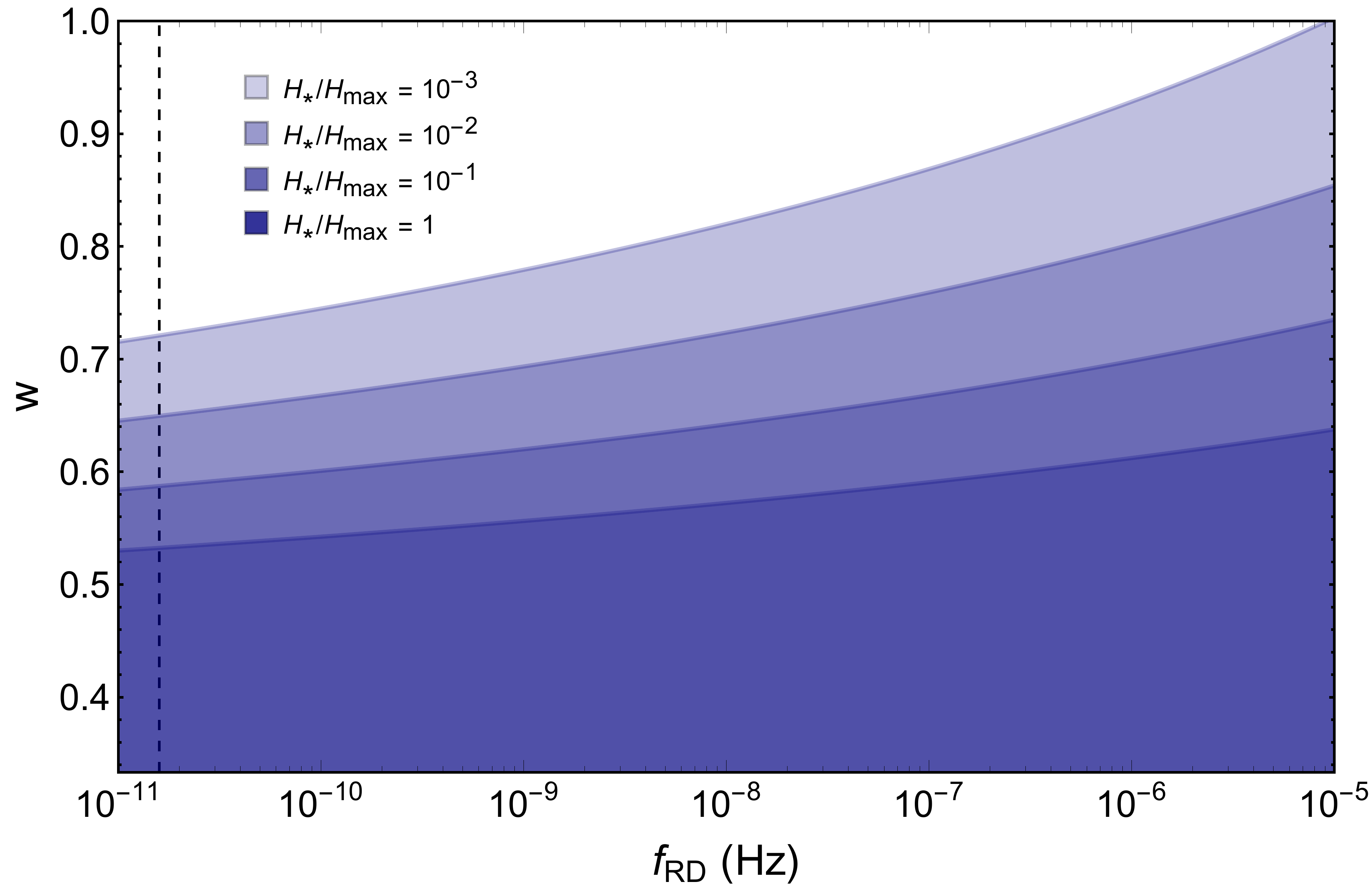}\vspace*{0.5cm}
	\includegraphics[width=10cm]{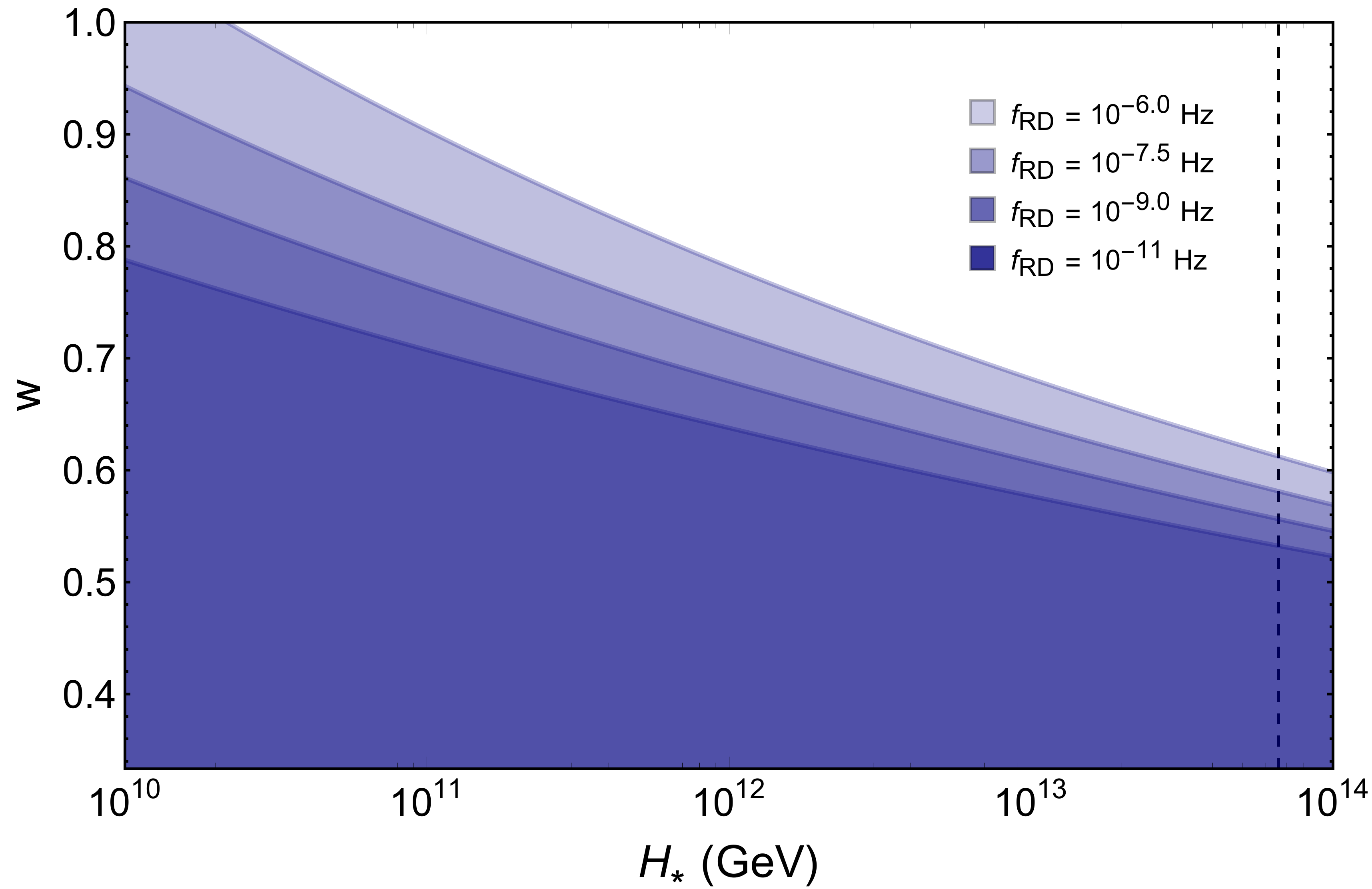}
	\caption{Colored regions represent the parameter space regions compatible with the BBN constraint, Eq.~(\ref{eq:BBNlocal}). The upper boundaries of these regions correspond to the function $w_{\rm BBN}(H_*,f_{\rm RD})$. In the top panel we show the regions compatible with the BBN bound in the $\lbrace \bar w,f_{\rm RD}\rbrace$ plane for a set of fixed values of $H_*$; the dotted line indicates the lowest value of $f_{\rm RD}$ for the Universe to reheat before the onset of BBN. In the bottom panel we show analogous regions in the $\lbrace \bar w,H_*\rbrace$ plane for a set of fixed values of $f_{\rm RD}$; the dotted line indicates the current upper bound on $H_*$ from CMB.}
	\label{fig:2}
\end{figure}

The remarkable fact that the critical value $\delta_c$ in Eq.~\eqref{eq:deltaCritical} does not depend on $H_*$ stems from the fact that the high frequency end of the GW energy spectrum $\Omega_{\rm GW}(t_0,k_*)$, which determines whether the BBN constraint is satisfied, is independent of $H_*$. This follows simply from Eq.~(\ref{eq:deltaGW}), where it is manifest that the energy density of GWs scales exactly as radiation, and furthermore that both the tensor spectrum and the initial energy of the radiation sector exhibit the same dependence on the Hubble scale $\propto (H_*/m_p)^2$. Since by construction $\Delta(t_0) \equiv \Omega_{\rm rad}^{(0)}$, we can obtain today's GW energy spectrum of the mode $k_*$ from Eq.~(\ref{eq:deltaGW}), simply as $\Omega_{\rm GW}(k_*,t_0) \sim {25\mathcal{A}_\alpha\over \pi^2\delta}\Omega_{\rm rad}^{(0)}$. Therefore, $\Omega_{\rm GW}(t_0,k_*)$ depends only on $\delta$ and $\alpha$. When we impose the BBN bound we obtain a constraint on $\delta$ that depends only mildly on $\alpha$ (as $\mathcal{A}_\alpha$ changes only slightly with $\alpha$). This leads to a GW spectrum that always ends on the same end point (modulo a mild dependence on $\alpha$), independently of $H_*$ and $\bar w$, see the top panel of Fig.~\ref{fig:3}. Only if we change $\delta$, we can appreciate a change (linearly inverse with $\delta$) in the amplitude of the high frequency point of the GW spectrum, see the bottom panel of Fig.~\ref{fig:3}.

\begin{figure}
	\center
	\includegraphics[width=10cm]{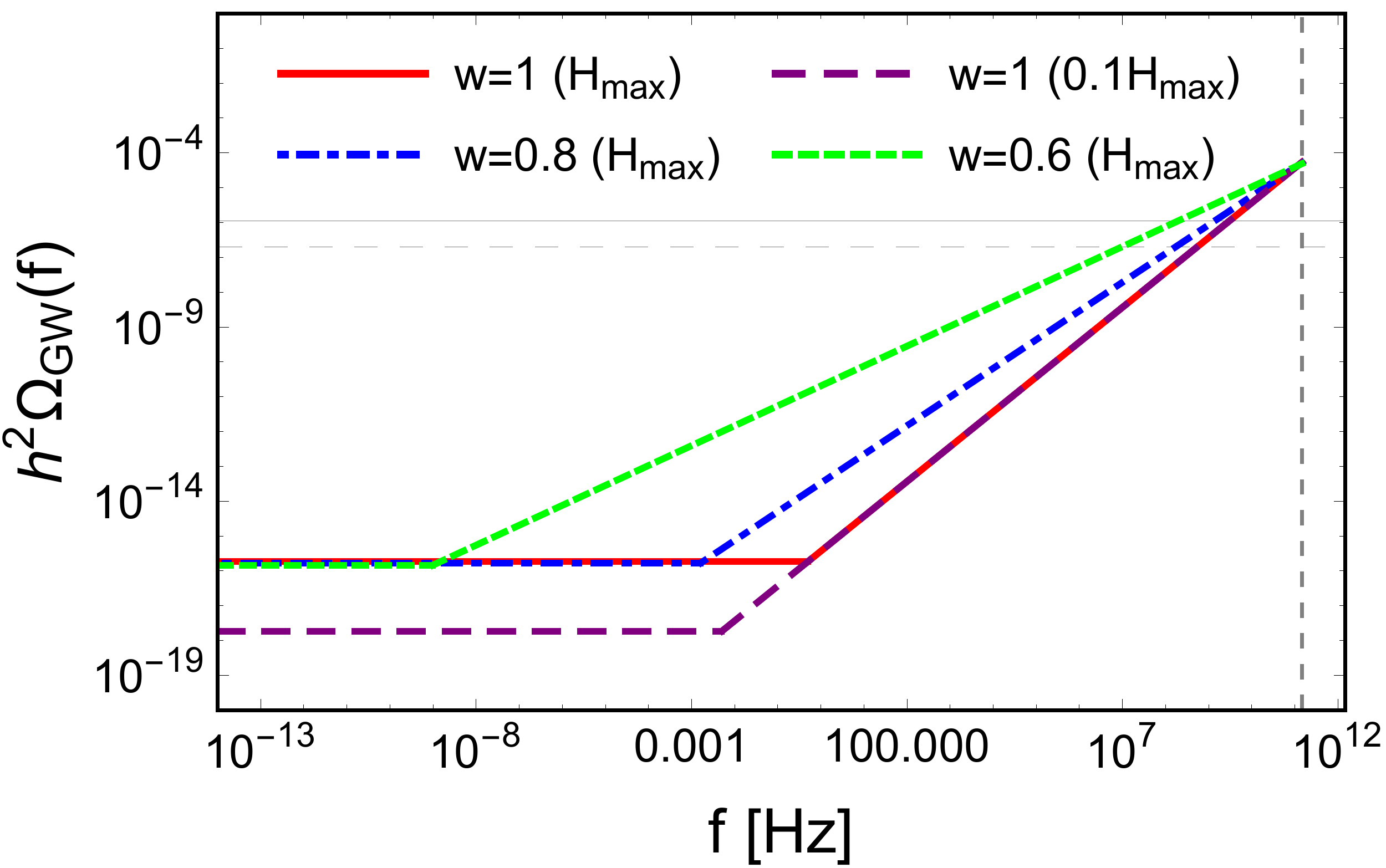}\vspace*{0.5cm}
	\includegraphics[width=10cm]{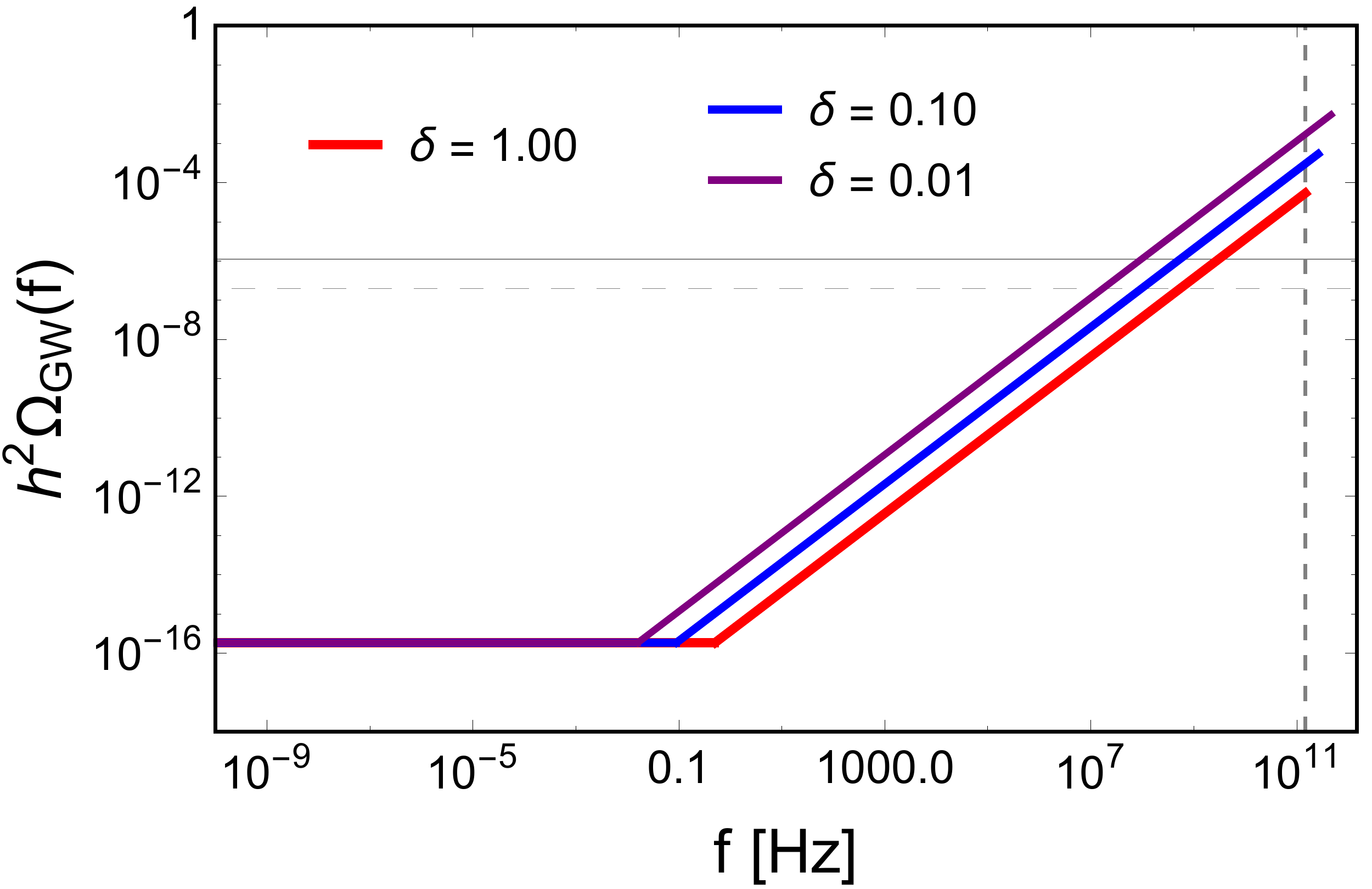}
	\caption{Different plots of the GW energy density spectrum today $h^2\Omega_{\rm GW}(f)$. In the top figure we fix $\delta = 1$, and plot  spectra for $H_* = H_{\rm max}$ and $\bar w \simeq 1$ (solid, red), $\bar w = 0.8$ (dotted-dashed, blue) and $\bar = 0.6$ (short-dashed, green), and also for $H_* = 0.1\cdot H_{\rm max}$ and $\bar w = 1$ (long-dashed, purple). In the bottom figure we fix $H_* = H_{\rm max}$ and $\bar w \simeq 1$, and plot spectra for $\delta = 1$ (red), $\delta = 0.1$ (blue), and $\delta = 1$ (purple). In both top and bottom figures, we indicate by horizontal lines the BBN (solid line) and CMB (dashed line) bounds on stochastic GW backgrounds, whereas the vertical dashed line indicates $f_{*}$ for $\delta = 1$.}
	\label{fig:3}
\end{figure}

In Fig.~\ref{fig:4}, we show the region in the parameter space $\lbrace \bar w, f_{\rm RD}\rbrace$ compatible with the BBN constraint for $H_* = 10^{13}$ GeV, i.e.~$\bar w \leq w_{\rm BBN}(f_{\rm RD}, H_* = 10^{13}~{\rm GeV})$. In the same figure, we also plot the corresponding values of $\bar w$ as a function of $f_{\rm RD}$ for $H_* = 10^{13}$ GeV and three values of $\delta$, namely $\delta = 0.01, 1, 48$. As we can see, the curves of $\bar w$ as a function of $f_{\rm RD}$ always lie above the coloured region for the generic values $\delta \leq 1$. Since the critical value $\delta_c$ in Eq.~(\ref{eq:deltaCritical}) does not depend on $H_*$, analogous plots for different values of $H_*$ exhibit trajectories of $\bar w$ as a function of $f_{\rm RD}$ lying always above the coloured regions for $\delta \leq 1$, independently of $H_*$. Only if $\delta \gtrsim 50$, we can then guarantee that the GW spectrum does not violate the BBN constraint. 

\begin{figure}
	\center
	\includegraphics[width=10cm]{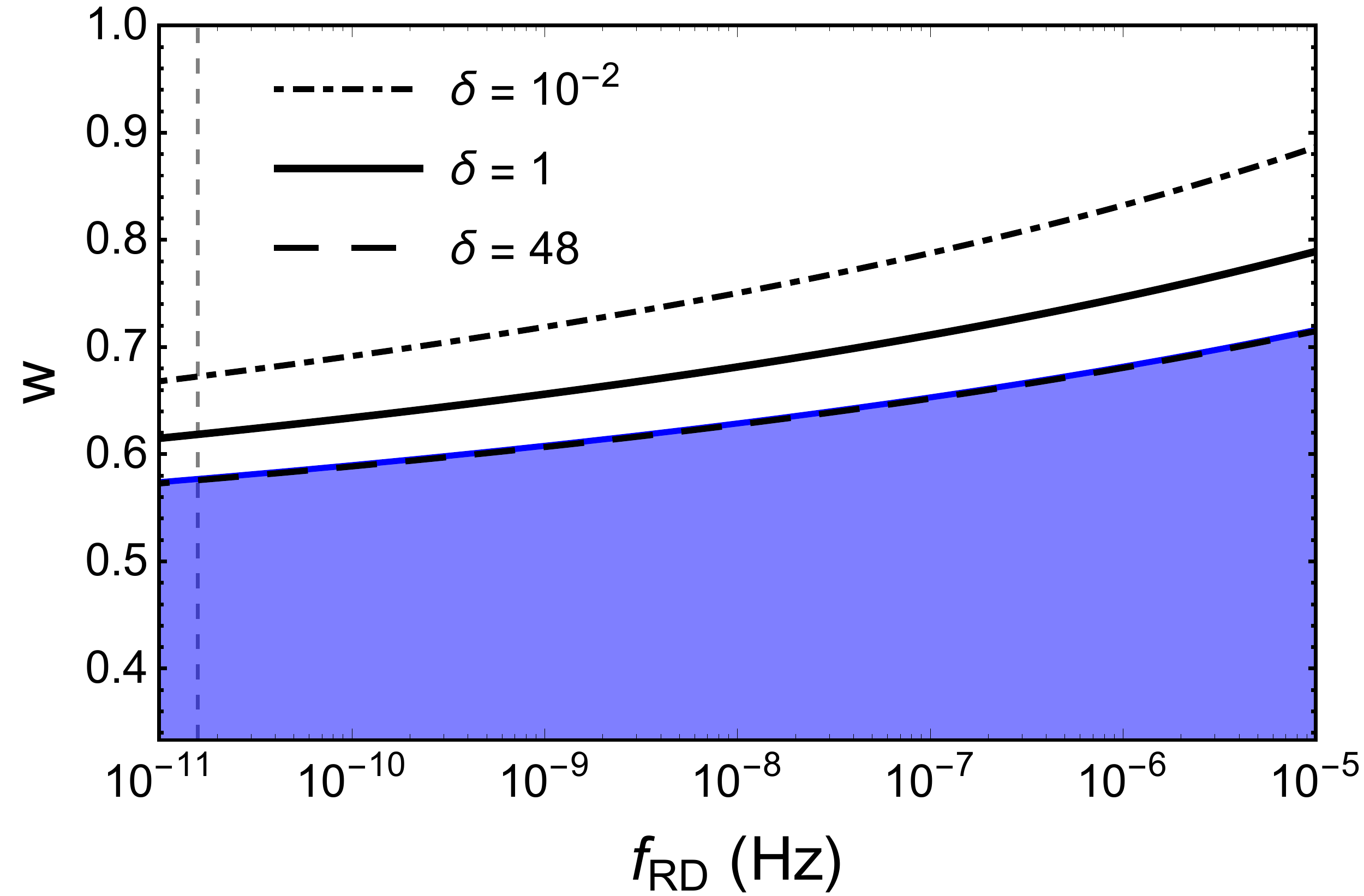}
	\caption{The coloured region corresponds to $\bar w \leq {w}_{\rm BBN}(f_{\rm RD},H_*)$, for $H_* = 10^{13}$ GeV. Also plotted are the curves of ${\bar w}$ as a function of $f_{\rm RD}$, for fixed $H_* = 10^{13}$ GeV, and $\delta = 0.01, 1, 48$ (dotted-dashed, solid and dashed lines, respectively). The vertical dashed line marks $f_{\rm RD} = f_{\rm BBN}$. For the generic values $\delta \leq 1$, the curves of $\bar w$ as a function of $f_{\rm RD}$ always lie above the coloured region, i.e. there is no solution in the parameter space compatible with the constraint~(\ref{eq:BBNlocal}). Only when $\delta \gg 1$, the GW spectrum may respect the BBN bound. Since the critical value $\delta \geq \delta_c$ does not depend on $H_*$, c.f.~Eq.~(\ref{eq:deltaCritical}), analogous plots for different values of $H_*$ always exhibit the same fact: in order to make the GW abundance compatible with the BBN constraint we need $\delta \gtrsim 40-50$.}
	\label{fig:4}
\end{figure}

Taken at face value, with the current constraint $\Delta N_\nu \lesssim 0.2$, the upper limit in the $rhs$ of the BBN constraint Eq.~(\ref{eq:BBNlocal}) corresponds to a situation where $\sim 90\%$ of the universe energy budget during RD correspond to photons and neutrinos while the other $\sim 10\%$ to gravitons, which is still a significant fraction. Hence, it is more realistic to demand that $h^2\Omega_{\rm GW}^{(0)} \ll 10^{-6}$. For instance, simply by considering that at the time of BBN the energy density of GWs does not represent more than say $\sim 1\%$ of the energy density of photons, the constraint becomes $h^2\Omega_{\rm GW}^{(0)} \lesssim 10^{-7}$ today, which, via Eq.~(\ref{eq:deltaCritical}) easily cranks up the minimum $\delta$ up to $\delta_{c} \sim 500$. The problem with this reasoning is that there is no clear value below the $rhs$ amplitude given in Eq.~(\ref{eq:BBNlocal}), to be used as an upper bound. A more precise argument can be obtained however from the Hubble rate at the time of CMB decoupling, as this can be also used to infer an upper bound on extra radiation components parametrized by $\Delta N_{\nu}$~\cite{Smith:2006nk,Sendra:2012wh,Pagano:2015hma}. This translates to an upper bound on the amount of GWs, which actually extends to a greater frequency range than the BBN bound, down to $f \lesssim 10^{-15}$ Hz~\cite{Smith:2006nk}. From such analysis, Ref.~\cite{Caprini:2018mtu} estimates the constraint implied by the Planck satellite (under the hypothesis of GWs with homogeneous initial conditions) as $h^2\Omega_{\rm GW}^{(0)} \lesssim 2\cdot 10^{-7}$. Substituting this value in the $rhs$ of Eq.~(\ref{eq:deltaCondition}), the critical value from Eq.~(\ref{eq:deltaCritical}) transforms to
\begin{align}\label{eq:deltaCritical}
\delta \geq \delta_c \equiv 83.9\cdot2^{\alpha}\mathcal{A}_\alpha \gg 1
\,.
\end{align}
For an EoS $\bar w \simeq 1$, we obtain $\delta_c \simeq 213$. In other words, using the CMB constraint we conclude that we need $\delta > \delta_c \sim 200$ in order to have a consistent scheme. This reinforces even further the idea that gravitational reheating with generic values $\delta \leq 1$ is inconsistent. 

The critical value of $\delta$ in Eq.~\eqref{eq:deltaCritical}, derived from the CMB constraint, is larger than those derived from the BBN constraint and the need that GWs must not dominate the radiation sector, simply because it is based on the demand that $\rho_{\rm GW}/\rho_{\rm rad} \lesssim \mathcal{O}(0.01)$, whereas BBN requires (given the current constraints on $\Delta N_\nu$) that $\rho_{\rm GW}/\rho_{\rm rad} \lesssim \mathcal{O}(0.1)$ and the need to have less GWs than other radiation is equivalent to $\rho_{\rm GW}/\rho_{\rm rad} \lesssim \mathcal{O}(1)$. Thus, the respective critical threshold in each case simply responds to the requisite that the GW background is not larger than a given required fraction of the radiation energy density. As the BBN/CMB constraints demand larger thresholds of $\delta$, we take those as a reference, and conclude that we need $\delta \gtrsim \mathcal{O}(100)$ in order to avoid any inconsistency with current observational bounds.

\section{Discussion and Conclusion}
\label{sec:Discussion}

We have arrived at the striking result that whenever the Universe seems to reheat successfully via gravitational reheating from the point of view of having RD before the onset of BBN, i.e. $T_{\rm RD} > T_{\rm BBN}$, the whole scenario is inconsistent unless very special set-ups are considered: the energy density of the high-frequency end of the resulting GW spectrum would simply be too large, violating various bounds on the presence of a stochastic GW background. The consistency condition to reheat into a radiation dominated Universe that is not dominated by GWs requires $\delta \gtrsim 5$. The consistency condition from BBN requires a stronger constraint $\delta \gtrsim 50$, and finally the consistency condition from CMB requires an even more stringent constraint $\delta \gtrsim 200$. This implies that $\delta$ must lie well outside the generic range $\delta\lesssim 1$, independently of the values of $H_*$ and $\bar w$.  

As we argued in Section~\ref{sec:MinimalRequisites}, in order to have $\delta\gg 1$ in standard gravitational reheating, some very special circumstance is needed. For instance, if there were $\mathcal{O}(100)$ species present during inflation, by adding up their energies we can reduce (linearly in the number of fields) the otherwise too large ratio $1/\Delta_* \gg 1$ of the inflationary energy density to that of the initial radiation. This allows the total energy of the radiation component to be sufficiently large compared to the energy stored in the GWs. By identifying $\delta \propto \mathcal{N} \sim \mathcal{O}(100)$ with the number of species, we obtain $\Omega_{\rm GW}(k_*,t)/\Delta(t) \sim 1/\mathcal{N} \sim \mathcal{O}(10^{-2})$, so that there would be no conflict with any upper bound on stochastic GW backgrounds. The problem, however, is that requiring the existence of said hundreds of fields is contrived, as the coupling strengths of these fields need to lie in very limited ranges. In particular, potential candidates for such fields include self-interacting field species with self-coupling tuned to $\lambda \sim 10^{-14}$. While the choice of such coupling value is not necessarily a tuning in itself, the presence of a large number of fields, e.g.~$\sim 300$, all with identically tuned coupling values, is unlikely. Non-conformal non-minimally coupled scalar fields with coupling to curvature $|1-6\xi| \sim 1$ are also viable candidates. We can exclude however the coupling values $\xi \simeq 1/6$ with $|1-6\xi| \ll 1$, because then the initial energy ratio of inflaton to radiation would be enhanced as $1/\Delta_* \propto 1/|1-6\xi|^2$, and hence the number of species needed would increase as $\mathcal{N} \gtrsim \mathcal{O}(10^2)/|1-6\xi|^2$. Thus, consistent gravitational reheating requires a very specific setup: that there exist (at least) $\mathcal{O}(10^2)$ fields non-minimally coupled to gravity but with non-conformal coupling in the window $|1-6\xi| \sim 1$. This appears contrived. On top of that, the origin of the primordial curvature perturbation in this case, must actually be provided by the inflationary sector. 

Some of the choices we made in deriving the bounds on $\delta$ are on the conservative side, i.e.~they yield less tension with the observational bounds. Changes in the number of relativistic $dof$ occurring in the stiff epoch does not affect the expansion rate and so does not affect the GW energy spectrum. To be on the conservative side, we have assumed that significant changes in the relativistic $dof$ occur in the radiation epoch. For instance, if the QCD phase transition occurs during the stiff era instead of in the radiation epoch, then it will  not contribute to lowering the GW energy spectrum, and the resulting high-frequency part of the GW energy spectrum would increase by roughly a factor of $\sim 2$ compared to the value we reported, hence increasing $\delta_c$ by the same factor. Furthermore, the estimated energy density of a non-minimally coupled field at the end of inflation as $\sim 0.01 H_* ^4$, c.f.~Eqs.~(\ref{eq:EnergyParam}) and (\ref{eq:InitialRatio}), is actually an upper bound, implicitly assuming that the transition from the (quasi-) de Sitter inflationary to the following decelerating regime occurs very rapidly (within less than a Hubble time). That is a model dependent assumption which in general does not need to be true. If the transition occurs more gradually, the amount of initial energy in the radiation component is further suppressed, effectively lowering the prefactor in Eq.~\eqref{eq:InitialRatio}, which means that in general a higher value of $\delta$ is needed.

In summary, even though -- strictly speaking -- there are setups where one can achieve $\delta \gg 1$, they seem artificial or $ad$-$hoc$. In general, if there is one species (i.e.~$\delta \leq 1$) excited during or towards the end of inflation, then the ratio of the energy it acquires by the end of inflation to the energy of the inflationary sector at that time, is simply too small. Thus, even though there exist solutions seemingly compatible with BBN (i.e.~$w_{\rm RD} \leq \bar w < 1$) so that the universe reheats successfully at a sufficiently large temperature, the amount of the enhanced GWs from inflation is too large compared to the energy of the radiation fields, making the whole set-up of gravitational reheating inconsistent.

Possible ways to reduce the amount of tension include considering a slight red-tilt in the inflationary GW spectrum. As noted below Eq.~(\ref{eq:consistencyCondition}), the current upper bound on the tensor-to-scalar ratio, implies already that the tilt is so small that (in the absence of running) the GW spectrum decays only by a factor $\sim 0.6$, when we compare the amplitude of the modes leaving the Hubble radius at the end of inflation and exiting the Hubble radius $\sim$ 50-60 e-folds before. Therefore, this would only slightly alleviate the critical thresholds derived above, reducing them by a factor $\delta_c \rightarrow 0.6\,\delta_c$. Thus, consistency of the CMB bound would still require, roughly speaking, $\delta \gtrsim 100$. Of course, in realistic inflationary models, typically the deviation from slow-roll becomes more noticeable towards the end of inflation, and this is expected to induce a running in the tensor tilt. This will reduce further the amplitude of the GW modes in the high frequency end  of the spectrum, but then it becomes a model dependent computation. For standard single field inflation monomial potentials, one typically obtains a reduction of the spectral energy amplitude at the highest frequency mode of the order of $\sim \mathcal{O}(0.1)$. In this case, the CMB constraint on stochastic GW backgrounds would translate into $\delta \gtrsim \mathcal{O}(10)$. This of course alleviates the tension as now only a reduced number of fields with tuned couplings are required. However, as said, this is a model dependent consideration and, in any case, it still requires an ad-hoc  scenario with at least dozens of fields having tuned couplings.

Gravitational reheating has been considered for years as a valid mechanism to reheat the universe without invoking direct couplings between the inflationary sector and other matter fields. The inconsistency we uncovered suggests however that the inflaton field must (most likely) be coupled to other field species, in order to transfer its energy into the latter so that the Universe can become radiation dominated. We reiterate here that this conclusion applies only to standard gravitational reheating as described in Section~\ref{sec:GravRHidea}. It is worth noticing that reheating in inflationary models based on modified gravity scenarios beyond General Relativity, such as Starobinsky inflation~\cite{Starobinsky:1980te, DeFelice:2010aj}, may actually proceed efficiently without the need to introduce a stiff epoch, as gravitationally coupled matter fields are automatically coupled to the inflationary sector driven by gravitational $dof$~\cite{zeldovich1977rate,starobinskii1981evolution}. In scenarios where the inflaton is non-minimally coupled to gravity with $f(\phi)R$, couplings between the inflaton and other matter fields can also arise spontaneously when the inflaton oscillates after inflation around the minimum of its potential~\cite{Watanabe:2006ku,Watanabe:2010vy}. Therefore, to reheat the universe in modified gravity inflationary models or in non-minimally coupled inflaton scenarios, there is no need for a stiff period following inflation, and hence the inconsistency uncovered in this paper does not apply in these cases. 

Furthermore, in the context of General Relativity, setups different than (though similar in spirit to) standard gravitational reheating, have been also envisaged to reheat the Universe. In particular, if the inflationary sector is not coupled directly to matter fields, but there is a spectator field during inflation with large non-minimal coupling $\xi \gtrsim 1$ and self-interacting potential, i.e.~$\xi\phi^2R + \lambda \phi^4$, the field develops a tachyonic instability if there is a stiff epoch after inflation. As the potential energy grows due to the instability, the self-interaction eventually compensates the tachyonic mass~\cite{Figueroa:2016dsc,Dimopoulos:2018wfg}. %From then on the field starts oscillating (anharmonically) with positive square mass around the quartic minimum of its potential. Due to the tachyonic growth, 
The fraction of energy to the inflationary sector at that moment is significantly enlarged compared to standard gravitational reheating, leading to an effective energy ratio with $\delta \sim 10^3 (\xi^2/\lambda) \gg 1$ (c.f.~Eq.~(\ref{eq:InitialRatio}) in this paper, and Eq.~(10) in Ref.~\cite{Figueroa:2016dsc}). This ratio naturally evades the problem we have uncovered in standard gravitational reheating. In fact, if we assume stability of the SM all the way up to the inflationary energy scale, a natural candidate for this reheating mechanism is actually the SM Higgs, see~\cite{Figueroa:2015rqa,Figueroa:2016dsc}. 

To conclude, the inconsistency just derived diminishes to some extent, the motivation for introducing in the first place an early stiff period after inflation. If the inflaton has direct couplings to other species, there seems to be no motivation to require such a stiff period. It is nonetheless possible to have setups where the inflaton is coupled to other species which it will eventually perturbatively decay into, or where the particle species are rapidly excited, a la {\it instant preheating}, during a stiff period~\cite{Felder:1999pv,Dimopoulos:2017tud,Haro:2018jtb}. The inconsistency uncovered in our work does not rule out, therefore, the existence of a stiff period after inflation, but it reduces the motivation for it. The same logic applies to the blue tilted GW background expected due to a stiff era.  The boost of the spectral amplitude at the high frequency end of the inflationary GW spectrum, makes this signal one of the most appealing backgrounds from the early Universe that can be potentially observed by direct detection GW experiments~\cite{Giovannini:1998bp,Giovannini:1999bh,Riazuelo:2000fc,Sahni:2001qp,Tashiro:2003qp,Boyle:2007zx,Giovannini:2008zg,Giovannini:2008tm,Artymowski:2017pua,Caprini:2018mtu}. As our result reduces however the motivation to introduce a stiff era, one could argue that the inflationary GW background is also expected, most likely, to remain as a (quasi-)scale invariant background, with an amplitude too small to be observable by direct detection experiments such as aLIGO or LISA.

\vspace*{0.5cm}

{\it Acknowledgements}. We are very grateful to M.~Shaposhnikov and S.~Sibiryakov for comments on the draft. The work of DGF was supported partially by the ERC-AdG-2015 grant 694896 and partially by the Swiss National Science Foundation (SNSF).

%=====================================================================
%=====================================================================
%=====================================================================

\appendix
\section{High-frequency behaviour of the GW energy spectrum in the presence of a stiff epoch }\label{appendix:GWspectrumderivation}

In order to derive the GW energy spectrum, we need to solve for the evolution of the tensor perturbation $h_{ij}$. For that purpose, it is convenient to Fourier- and polarization-mode decompose the tensor perturbation as follows
\begin{equation}
h_{ij}(t,\mathbf{x})=\sum_{\lambda} \int \frac{d^3\mathbf{k}}{(2\pi)^3}h_{\mathbf{k}}^{\lambda}(t)e^{i\mathbf{k}.\mathbf{x}}\epsilon_{ij}^\lambda(\mathbf{k})\,,
\end{equation}
where $\epsilon^\lambda_{ij}(\mathbf{k})$ are the polarization tensors and $\lambda$ stands for $+$ or $\times$ polarization states. In what follows, we will assume that the GW spectrum is unpolarized $\left<|h_{\mathbf{k}}^+|\right>=\left<|h_{\mathbf{k}}^+|\right>$ and isotropic $h_{\mathbf{k}}^\lambda=h_{k}^\lambda$,  with $k\equiv |\mathbf{k}|$. Hence, we will drop the polarization indices $\lambda$ and simply write the tensor mode function as $h_k$, which is related to the GW power spectrum as
\begin{equation} \label{eq:powerspectrummodefunction}
\Delta_h^2(t,k)=\frac{k^3}{2\pi^2}4\left<|h_k(t)|^2\right>\,.
\end{equation}
Deep inside the stiff epoch when the EoS $\bar w$ is essentially constant, and in the absence of anisotropic stress-tensor perturbation, $h_k$ obeys the equation of motion 
\begin{equation}
h_k''+\frac{2\alpha}{t}h_k'+k^2h_k=0\,,
\end{equation}
where primes denote derivatives with respect to the conformal time $t$. The solution to the above equation for modes that cross the horizon during the stiff epoch is \cite{Boyle:2005se}
\begin{equation}\label{eq:stiffsol}
h_{k\gg k_{\rm RD}}(t<t_{\text{RD}})=\Gamma\left(\alpha+\frac{1}{2}\right)\left(\frac{2}{kt}\right)^{\alpha-\frac{1}{2}}J_{\alpha-\frac{1}{2}}(kt)h_k^{\rm inf}\,,
\end{equation}
where $J_\nu(x)$ is the Bessel function of the first kind. This solution is obtained by imposing the boundary condition that the mode function must match with the value predicted by inflation $h_k^{\rm inf}$ in the superhorizon limit $kt\ll 1$ (plugging $h_k^{\rm inf}$ into Eq.~\eqref{eq:powerspectrummodefunction} gives us $\Delta_{h,\rm inf}^2(k)$ introduced in Eq.~\eqref{eq:InfSpectrum}).

The oscillation-average of the square of the sub-horizon limit of Eq.~\eqref{eq:stiffsol} is
\begin{align}\label{eq:stiffsolsquareavg}
\overbar{\left|h_{k\gg k_{\rm RD}}(t<t_{\text{RD}})\right|^2}=&\frac{\Gamma^2\left(\alpha+\frac{1}{2}\right)}{2\pi}\left(\frac{2}{kt}\right)^{2\alpha}|h_k^{\rm inf}|^2\,.
\end{align}
Combining Eq.~\eqref{eq:aRd2aBBN} and the $t\gg (a_*H_*)^{-1}$ limit of Eq.~\eqref{eq:stiffscalefactor} , we get
\begin{equation}\label{eq:atlimitrelation}
a(t)=\left(\frac{1}{\sqrt{2}\alpha}\right)^\alpha k_{\rm RD}^\alpha a_{\rm RD}t^\alpha,\quad (a_*H_*)^{-1}\ll t\ll t_{\rm RD}
\end{equation}
where we have also used $k_{\rm RD}=a_{\rm RD} H_{\rm RD}$ in the process. The above relation, with the aid of $k_{\rm RD}/k=f_{\rm RD}/f(k)$, allows us to rewrite Eq.~\eqref{eq:stiffsolsquareavg} as
\begin{align}
\overbar{\left|h_{k\gg k_{\rm RD}}(a)\right|^2}=\frac{\Gamma^2\left(\alpha+\frac{1}{2}\right)}{2\pi}\left(\frac{\sqrt{2}f_{\rm RD}}{\alpha f(k)}\right)^{2\alpha}\left(\frac{a_{\rm RD}}{a}\right)^2|h_k^{\rm inf}|^2\,.
\end{align}
Since sub-horizon tensor perturbations scale as $h_k\propto a^{-1}$, the above expression, though obtained by solving the GW equation of motion in the stiff epoch, is also valid in all the subsequent epochs. Plugging the above equation into Eq.~\eqref{eq:powerspectrummodefunction} and using Eq.~\eqref{eq:InfSpectrum}, we find
\begin{align}
\Delta_h^2\left(t,f\gg f_{\rm RD}\right)=\frac{\Gamma^2\left(\alpha+\frac{1}{2}\right)}{\pi^3}\left(\frac{2}{\alpha^2}\right)^\alpha\left(\frac{a_{\rm RD}}{a(t)}\right)^2\left(\frac{f_{\rm RD}}{ f}\right)^{2\alpha}\left(\frac{H_*}{m_p}\right)^2\,,
\end{align}
which can then be inserted into Eq.~\eqref{eq:GWenergySpectrum} to give us the high-frequency part of the present-day the GW energy spectrum corresponding to modes that cross the horizon during the stiff epoch
\begin{align}\label{eq:GWhighfreqenergyspectrum}
\Omega_{\rm GW}\left(t_0,f\gg f_{\rm RD}\right)=\frac{\Gamma^2\left(\alpha+\frac{1}{2}\right)}{6\pi^3}\left(\frac{2}{\alpha^2}\right)^\alpha\Omega_{\rm rad}^{(0)}\left(g_{s,0}\over g_{s,k(f)}\right)^{4\over3}\nonumber\left(g_{*,k(f)}\over g_{*,0}\right)\left(\frac{f}{f_{\rm RD}}\right)^{2-2\alpha}\left(\frac{H_*}{m_p}\right)^2\,.
\end{align}
In arriving at the above form, we used Eqs.~\eqref{eq:aRd2a0}, $k_{\rm RD}/k(f)=f_{\rm RD}/f$, and $k_{\rm RD}=a_{\rm RD}H_{\rm RD}$. The scale-invariant part of the GW energy spectrum corresponding the modes that cross the horizon during the RD epoch can be obtained by setting $\alpha=1/2$, i.e. $\bar{w}=1/3$, which amounts to considering the standard cosmological scenario without a stiff epoch. Doing so gives us Eq.~\eqref{eq:InfGWtodayRD}. We can then rewrite Eq.~\eqref{eq:GWhighfreqenergyspectrum} as
\begin{align} 
\Omega_{\rm GW}\left(t_0,f\gg f_{\rm RD}\right)={\left.\Omega_{\rm GW}^{(0)}\right|_{\rm RD}\Gamma^2(\alpha+{1\over2})\over2^{(1-\alpha)}\alpha^{2\alpha}\Gamma^2({3\over2})} \left(\frac{f}{f_{\rm RD}}\right)^{2-2\alpha}\,.
\end{align}
Note that $\Gamma(3/2)=\sqrt{\pi}/2$. 

In this derivation we have assumed that the transition from the stiff to radiation epoch occurs smoothly due to the difference in the scaling law of the energy density of radiation ($\propto a^{-4}$) and stiff fluid $\propto a^{-3(1+w)}$. The full GW energy spectrum in this case cannot be solved analytically, but we have managed to extract the high-frequency behaviour of the spectrum corresponding to modes that cross the horizon long (a few Hubble times) before the RD epoch begins. For the sake of solubility, it is often assumed in the literature that the stiff-radiation transition happens instantaneously, i.e. much faster than the Hubble time at the moment of transition. The GW energy density obtained here is smaller by a factor of $2^{1-\alpha}$, which ranges from $1$ (if $\bar{w}=1/3$) to $2^{1/2}$ (if $\bar w=1$), compared to that obtained in the instantaneous transition case.

For a fully detailed derivation of the GW energy density spectrum in the case of both instantaneous and smooth transitions, see Ref.~\cite{Figueroa:2019paj}.

%\newpage
\vspace*{0.1cm}
\begin{center}
	--------------
\end{center}
%\newpage

\bibliography{StiffBiB}
\bibliographystyle{JHEP}

	\end{document}